\documentclass[aps,prb,reprint,showpacs,superscriptaddress,nofootinbib,floatfix]{revtex4-1}

\usepackage{color}
\usepackage{amsmath}
\usepackage{amssymb}
\usepackage{graphicx}
\usepackage{bm}

\newcommand{\HN}[1]{{#1}}

\begin{document}

\title{Nonequilibrium processes from Generalised Langevin Equations: \\
realistic nanoscale systems connected to two thermal baths 
}

\author{H. Ness}\email{herve.ness@kcl.ac.uk}
\affiliation{Department of Physics, Faculty of Natural and Mathematical Sciences,
King's College London, Strand, London WC2R 2LS, UK}

\author{A. Genina}
\affiliation{Department of Physics, Faculty of Natural and Mathematical Sciences,
King's College London, Strand, London WC2R 2LS, UK}

\author{L. Stella}
\affiliation{Atomistic Simulation Centre, School of Mathematics and Physics, Queen's
University Belfast, University Road, Belfast BT7 1NN, Northern Ireland,
UK}

\author{C.D. Lorenz}
\affiliation{Department of Physics, Faculty of Natural and Mathematical Sciences,
King's College London, Strand, London WC2R 2LS, UK}

\author{L. Kantorovich}
\affiliation{Department of Physics, Faculty of Natural and Mathematical Sciences,
King's College London, Strand, London WC2R 2LS, UK}

\begin{abstract}
We extend the Generalised Langevin Equation (GLE) method [Phys. Rev. B {\bf 89}, 134303 (2014)]
to model a central classical region connected to two realistic thermal baths at 
two different temperatures. 
In such nonequilibrium conditions a heat flow is established, via the central system, in
between the two baths.
The GLE-2B (GLE two baths) scheme permits us to have a realistic description of both the 
dissipative central system and its surrounding baths.
Following the original GLE approach, the extended Langevin dynamics scheme is modified
to take into account two sets of auxiliary degrees of freedom corresponding to the 
mapping of the vibrational properties of each bath. These auxiliary
variables are then used to solve the non-Markovian dissipative dynamics of the central
region. 
The resulting algorithm is used to study a model of a short Al nanowire connected 
to two baths. 
The results of the simulations using the GLE-2B approach are compared to the results
of other simulations that were carried out using standard thermostatting 
approaches (based on Markovian Langevin and Nose-Hoover thermostats).
\HN{
We concentrate on the steady state regime and study the establishment of a local
temperature profile within the system.
The conditions for obtaining a flat profile or a temperature gradient are
examined in detail, in agreement with earlier studies. 
The results show that the GLE-2B approach is able to treat, within a single scheme, 
two widely different thermal transport regimes, i.e. ballistic systems, with no 
temperature gradient, and diffusive systems with a temperature gradient.
}
\end{abstract}

\pacs{05.10.Gg, 05.70.Ln, 02.70.-c, 63.70.+h}

\maketitle

\section{Introduction}
\label{sec:intro}

Nanoscale devices and materials are becoming increasingly important in the development of novel 
technologies. In most applications of these new nanotechnologies, the central system is part 
of a more complex set up where driving forces are present to establish heat and/or particle flows. 
The understanding of the corresponding non-equilibrium properties is of utmost importance. 
This is  especially true when one considers potential applications based on the thermal 
conductivity of materials \cite{Berber2000,Kim2001,Shi2002,Padgett2004,Hu2008,%
Padgett2006,Yang2008,Estreicher2009}
and the heat transport within nanodevices 
\cite{Majumdar:1999,Segal2002,Segal:2003,Mingo:2003,Yao:2005,Wang:2007,%
Dubi2011,Widawsky:2012,Cahill2002,Pop2010,Zebarjadi2012,Gotsmann:2013,Menges:2013,Meier:2014,Li:2015}.

Being able to describe the dynamics and dissipation of such nanoscale atomic systems is central 
for modern nanoscience.
For that, one has to consider the central region of interest as an open system
surrounded by a heat bath (an environment) which is in contact with the system
and is kept at a given temperature. 
For studying the transport properties, one has to consider the proper non-equilibrium
conditions, i.e. the central region is connected to two (or more) independent heat
baths (kept at their own temperatures). Hence a heat flow (transient or stationary depending
on the experimental conditions) is established between the two baths via the central
region.
 
An appropriate general approach for treating this kind of systems is based
on the so-called Generalised Langevin Equation (GLE)\cite{Mori:1965,Adelman:1976,Adelman:1980,%
Ermak:1980,Carmeli:1983,Cortes:1985,%
Lindenberg:1990,Tsekov:1994a,Tsekov:1994b,Risken:1996,Hernandez:1999,Zwanzig:2001,%
Segal:2003,Kupferman:2004,Bao:2004,Luczka:2005,Izvekov:2006,Snook:2007,%
vanVliet:2008,Kantorovich:2008,Ceriotti:2009,Ceriotti:2010,Siegle:2010,%
Kawai:2011,Morrone:2011,Ceriotti:2011,%
Pagel:2013,Leimkuhler:2013,Baczewski:2013,Stella:2014,Ness:2015}.
The GLE is an equation of motion containing
non-Markovian stochastic processes where the particle (point particle with
mass) has a memory effect to its velocity.

The GLE has been derived for a realistic system of $N$ particles 
coupled to a single realistic (harmonic) bath, i.e. a bath described at the atomic 
level \cite{Kantorovich:2008}.
The non-Markovian dynamics is obtained for the central system with Gaussian 
distributed random forces and a memory kernel 
that is exactly proportional to the 
random force autocorrelation function \cite{Kantorovich:2008}.
Solving the GLE for complex heterogeneous and extended systems is still a challenge.
A major step towards the solution of this problem for a realistic application has been 
recently given 
\cite{Ceriotti:2009,Ceriotti:2010,Morrone:2011,Ceriotti:2011,Stella:2014}.
In particular, an efficient and transferable algorithm has been developed in 
Ref.~[\onlinecite{Stella:2014}] to solve the GLE numerically while taking into account 
the two fundamental features of the GLE $-$ a time-dependent memory kernel and the presence 
of a coloured noise, which are absolutely essential for the description of the bath at the 
atomic level.

However, the previous tools have been developed for a single bath only. In order to treat
properly the presence of a (transient or steady) flow of heat current through the central
system, one has to take into account the proper non-equilibrium conditions. 
That is, one has to consider the presence of at least two baths (at their own temperatures)
in contact with the central region.

The aim of the present paper is to extend the previous GLE approaches developed in
Refs.~[\onlinecite{Kantorovich:2008,Stella:2014,Ness:2015}] to the systems consisting
of a central region connected to two spatially separated thermal baths. 
With such an approach, we can study the non-equilibrium processes in nanoscale 
systems by using molecular dynamics (MD) simulations. 
The dissipative processes are properly described since the system can exchange energy (heat) 
with the environment. The environment consists of the two baths whose dynamical 
properties are described more thoroughly than when standard thermostats (i.e. Langevin, Nose-Hoover,
velocity renormalisation, etc) \cite{Andersen:1980,Nose:1984a,Nose:1984b,Hoover:1985,Toton:2010}
are used in conventional MD simulations.
Note that, for a system at equilibrium, the Langevin dynamics applied only to a part
of the system can be derived from the GLE assuming short-range atomic interactions
and the Markovian approximation \cite{Kantorovich:2008b}.
It was also shown in [\onlinecite{Kantorovich:2008b}] that thermostatting only some of the
degrees of freedom by using the Langevin dynamics brings the system, in equilibrium, to the
corresponding canonical distribution. 
It can also be shown that the application of the Nose method to only a part of the system
(i.e. to a subset of atomic degrees of freedom) also performs correct thermostatting of
the entire system to the target temperature. 

The paper is organised as follows.
In Sec.~\ref{sec:generalisation}, we generalise the methodology of 
Refs.~[\onlinecite{Kantorovich:2008,Stella:2014,Ness:2015}]
to the cases of two independent baths $\nu=1,2$ each having its temperature $T_\nu$.
The generalisation, called GLE-2B, includes the use of two sets of auxiliary degrees-of-freedom (DOF)
corresponding to each bath and their stochastic dynamics. 
This is performed by the use of a multi-variate Markovian
stochastic process for the position and momentum of the DOF of the central region and 
the corresponding ``position'' and ``momentum'' of the two sets of auxiliary DOF \cite{Stella:2014}.
The resulting algorithm is explicitly described in Appendix \ref{app:algo} 
and has been implemented in the code LAMMPS \cite{Plimpton:1995} following our previous
work on the GLE with a single bath \cite{Ness:2015}.
In Sec.~\ref{sec:gle_ex}, we consider some applications of the GLE-2B for a specific realistic system. 
It consists of a short Al nanowire connected to two Al baths. 
Each bath is represented by a set of auxiliary DOF generated from a model solid, i.e. one 
half-sphere of an Al fcc lattice. 
In this first application of our GLE-2B approach, we concentrate on the steady-state properties of
the system.
First, we consider the equilibrium condition and 
an artificially thermally decoupled system to perform a first
validation of our methodology. 
Then we treat the proper non-equilibrium conditions when $T_1 \ne T_2$.
We also compare our results with other possible thermostatting procedures.
In Sec.~\ref{sec:NE}  
we interpret our results for the temperature profiles through the system in terms of the properties of 
integrable versus non-integrable cases. 
In the latter case, a full temperature gradient is established in the system, while in the
former case there is no temperature gradient built up in the system. 
In terms of transport properties, a perfect thermal conductor (a ballistic thermal system) has
an infinite conductance (in the thermodynamic limit) and, therefore, there is no temperature gradient within
the central part of the system. Whenever the system presents some form of thermal resistance (finite
conductance), a temperature gradient exists in the system.
Finally, in Sec.~\ref{sec:ccl}, we conclude and discuss further developments of the present study.

\section{Generalisation to the two baths problem}
\label{sec:generalisation}

\begin{figure}
\begin{centering}
\includegraphics[width=70mm]{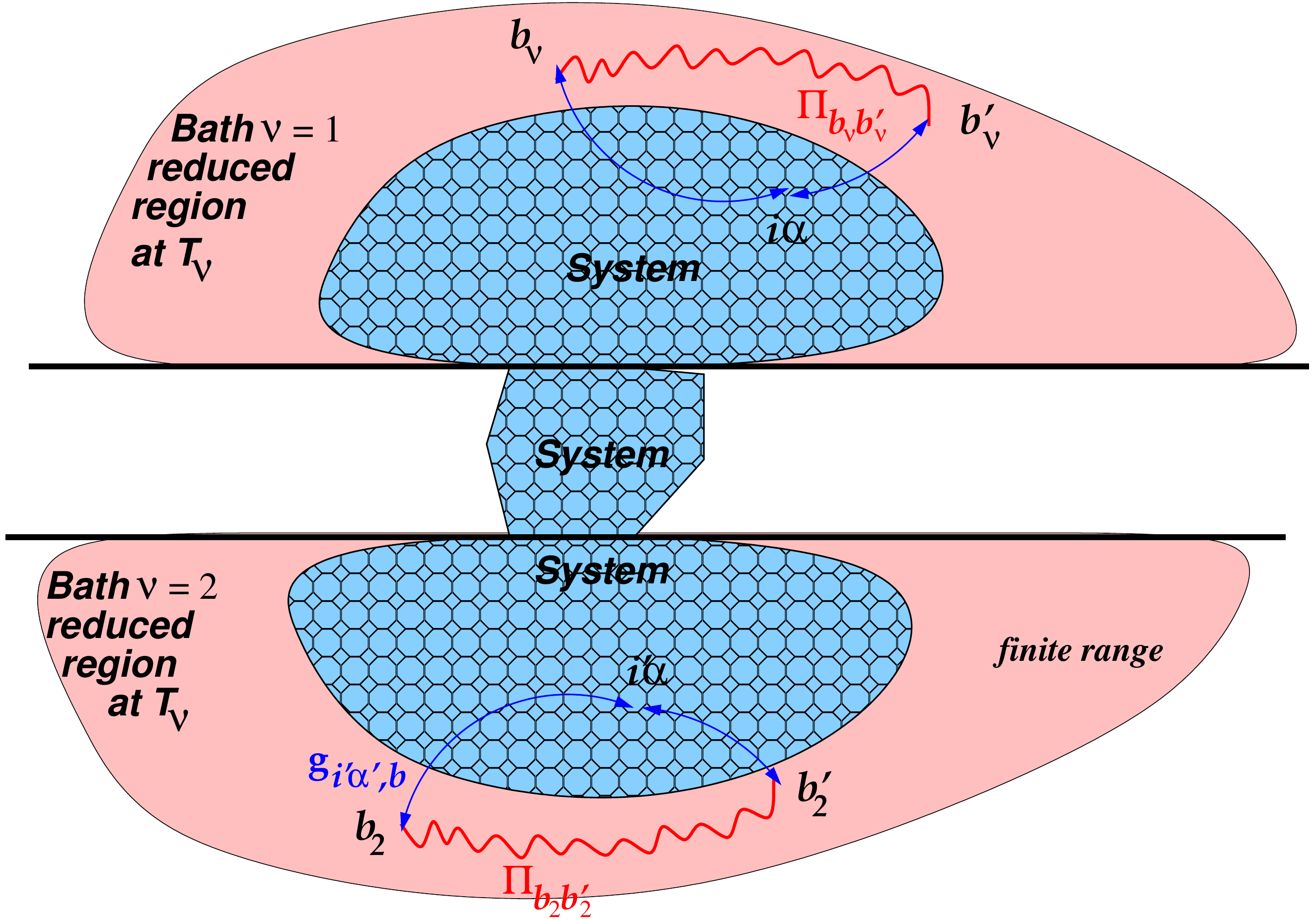}
\end{centering}
\caption{
(Colour online) Schematic representation of the system. 
It includes the finite size central system (in blue)
where the GLE dynamics is performed, and the two bath $\nu=1,2$ regions at temperature $T_\nu$. 
Because the forces $f_{b_\nu}$ and the quantities $g_{i\alpha,b_\nu}$ are of finite range (not
necessarily short ranged), one can perform the mapping of $\Pi_{b_\nu b'_\nu}(\omega)$ on a finite region
of space: the bath reduced region (in pink), one for each bath. }
\label{fig:sys}
\end{figure}

\subsection{Equations of motion for a system coupled to two baths: embedding Newton's equations}
\label{sec:GLEgeneralisation}

We consider a central system (of finite size) interacting with two independent baths $\nu=1,2$
[\onlinecite{Note:1}],
see Fig.~\ref{fig:sys} for a schematic representation of the system.
The corresponding classical Lagrangian is given by 
$\mathcal{L}\equiv\mathcal{L}_{{\rm sys}}+\mathcal{L}_{{\rm bath,(1)}}+\mathcal{L}_{{\rm bath,(2)}}+\mathcal{L}_{{\rm int}}$,
where
\begin{equation}
\mathcal{L}_{\rm sys}\left(\mathbf{r},\dot{\mathbf{r}}\right)=\sum_{i\alpha}\frac{1}{2}m_{i}\dot{r}_{i\alpha}^{2}-V\left({\bf r}\right)
\label{eq:lagrangian-1}
\end{equation}

\begin{equation}
\mathcal{L}_{{\rm bath},(1)}\left(\mathbf{u}_1,\dot{\mathbf{u}}_1\right)
=\sum_{l_1\gamma}\frac{1}{2}\mu_{l_1}\dot{u}_{l_1\gamma}^{2} -V^{\rm harm}_{(1)}(\mathbf{u}_1)
\label{eq:lagrangian-2}
\end{equation}

\begin{equation}
\mathcal{L}_{\rm int}\left(\mathbf{r},\mathbf{u}\right)=
\mathcal{L}_{{\rm int},(1)}\left(\mathbf{r},\mathbf{u}_1\right) + 
\mathcal{L}_{{\rm int},(2)}\left(\mathbf{r},\mathbf{u}_2\right) \ .
\label{eq:lagrangian-3}
\end{equation}
The positions of the atoms, labelled $i=1,2,...,N$ with mass $m_i$, of
the central system are given by vectors $\mathbf{r}$ with components $r_{i\alpha}$
($\alpha$ indicating the appropriate Cartesian coordinate). 
$\mathcal{L}_{\rm sys}$ is the Lagrangian
of the system with potential energy $V(\mathbf{r})$.
The Lagrangian $\mathcal{L}_{{\rm bath},(1)}$
describes the harmonic bath (bath $\nu=1$). The bath's atoms are labelled $l_1$ 
and have masses $\mu_{l_1}$. We introduce a shorthand notation for 
the labels of the bath degrees of freedom (DOF) $b_1 \equiv l_1 \gamma$, where $\gamma$ 
indicates the Cartesian coordinate. 
The corresponding potential energy $V^{\rm harm}_{(1)}(\mathbf{u}_1)$
is harmonic with respect to the displacements $u_{l_1\gamma}=u_{b_1}$
of the bath atoms from their equilibrium positions.
Its expression is 
\begin{equation}
V^{\rm harm}_{(1)}(\mathbf{u}_1) = \frac{1}{2} \sum_{b_1,b'_1}\sqrt{\mu_{l_1}\mu_{l'_1}}
u_{b_1}D^{(1)}_{b_1,b'_1}u_{b'_1}
\nonumber
\end{equation}
where $D^{(1)}_{b_1,b'_1}$ are the elements of the dynamic matrix of the bath $\nu=1$.
The Lagrangian $\mathcal{L}_{{\rm bath},(2)}$ for the bath $\nu=2$ is similar to 
$\mathcal{L}_{{\rm bath},(1)}$ and obtained from $\mathcal{L}_{{\rm bath},(1)}$
by swapping the bath index $1\leftrightarrow 2$.

Finally the interaction between the central system and the baths is a linear
superposition of the interaction between the system and each bath (we recall
that the baths are independent). The individual contribution 
$\mathcal{L}_{int,(\nu)}\left(\mathbf{r},\mathbf{u}_\nu\right)$ is taken
to be linear in the bath displacements $\mathbf{u}_\nu$ with the following
expression:
\begin{equation}
\mathcal{L}_{{\rm int},(\nu)}\left(\mathbf{r},\mathbf{u}_\nu\right) = 
-\sum_{b_\nu} \mu_{l_\nu }f_{b_\nu}(\mathbf{r}) u_{b_\nu} \ .
\label{eq:lagrangian-3bis}
\end{equation}
Note that the dependence of such an interaction on the system DOF, 
via $f_{b_\nu}(\mathbf{r})$, remains arbitrary.

We can now derive the equations of motion for the central system and baths DOF
from the Lagrangian, Eqs. (\ref{eq:lagrangian-1}-\ref{eq:lagrangian-3bis}),
following Refs.~[\onlinecite{Kantorovich:2008,Stella:2014}].
We find, for the central system DOF, that
\begin{equation}
m_{i}\ddot{r}_{i\alpha} = 
-\frac{\partial V\left(\mathbf{r}\right)}{\partial r_{i\alpha}}
-\sum_{\nu=1}^2 \sum_{b_\nu} \mu_{l_\nu} g_{i\alpha,b_\nu}(\mathbf{r}) u_{b_\nu}
\label{eq:sys_eoms}
\end{equation}
where $g_{i\alpha,b_\nu}(\mathbf{r})=\partial f_{b_\nu}({\bf r})/\partial r_{i\alpha}$.

For the bath DOF, we find two sets of equations which can be solved analytically, 
since the Lagrangian 
$\mathcal{L}_{{\rm bath},(\nu)}+\mathcal{L}_{{\rm int},(\nu)}\left(\mathbf{r},\mathbf{u}_\nu\right)$ 
is harmonic in the bath DOF $\mathbf{u}_\nu$. These sets of equations are given by
\begin{equation}
\mu_{l_\nu} \ddot{u}_{b_\nu} = 
-\sum_{b'_\nu} \sqrt{\mu_{l_\nu}\mu_{l'_\nu}}D^{(\nu)}_{b_\nu,b'_\nu}u_{b'_\nu}
-\mu_{l_\nu}f_{b_\nu}\left(\mathbf{r}\right)
\label{eq:bath_eom}
\end{equation}
for $\nu=1,2$. Equation (\ref{eq:bath_eom}) can be solved by introducing the kernel of the differential equation
defined from the eigenstates $v^{(\lambda)}_{b_\nu}$ and eigenvalues $\omega^2_{\nu,\lambda}$
of the dynamical matrix $D_{b_\nu,b'_\nu}$.
The solution of Eq.~(\ref{eq:bath_eom}) is then substituted into Eq.~(\ref{eq:sys_eoms}) to obtain 
a closed equation in terms of the system DOF only.

We consider the initial positions and velocities of the bath atoms, 
appearing in the solution of Eq.~(\ref{eq:bath_eom}), being stochastic.
It permits us to derive a generalised Langevin-like 
equation of motion (EOM) for the system DOF \cite{Kantorovich:2008,Stella:2014}:
\begin{multline}
m_{i}\ddot{r}_{i\alpha} = -\frac{\partial\bar{V}\left(\mathbf{r}\right)}{\partial r_{i\alpha}}\\
-\int_{-\infty}^{t}{\rm d}t' \sum_{\nu, i'\alpha'} K^{(\nu)}_{i\alpha,i'\alpha'}(t,t';\mathbf{r})
	\dot{r}_{i'\alpha'}(t')
+ \sum_\nu \eta^{(\nu)}_{i\alpha}(t;{\bf r}) .
\label{eq:gle}
\end{multline}
The dynamics of the system DOF is governed by an effective potential $\bar{V}$, two memory Kernels
$K^{(\nu)}_{i\alpha,i'\alpha'}(t,t';\mathbf{r})$ and stochastic forces
$\eta^{(\nu)}_{i\alpha}(t;{\bf r})$ corresponding to each independent bath $\nu$.

The potential energy $\bar V$ is given by the nominal potential energy $V$ inside the central system 
plus the potential energy between the central system and the two frozen baths. There is also
a ``polaronic'' correction energy due to the coupling between
the system atoms and the harmonic displacements of the baths'
atoms around their equilibrium positions:
\begin{equation}
\begin{split}
\bar{V}(\mathbf{r}) & = V(\mathbf{r}) 
-\frac{1}{2}\sum_\nu \sum_{b_\nu, b'_\nu} \sqrt{\mu_{l_\nu} \mu_{l'_\nu}}
f_{b_\nu}(\mathbf{r}) \Pi_{b_\nu b'_\nu}(0) f_{b'_\nu}(\mathbf{r}) \ .
\end{split}
\label{eq:effec_pot_matrix}
\end{equation}
The memory kernel for the bath $\nu$ is given by
\begin{equation}
\begin{split}
K^{(\nu)}_{i\alpha,i'\alpha'}(t,t';\mathbf{r})
=
\sum_{b_\nu, b'_\nu} \sqrt{\mu_{l_\nu}\mu_{l'_\nu}} g_{i\alpha,b_\nu}(\mathbf{r}(t)) \\
\Pi_{b_\nu,b'_\nu}(t-t') g_{i'\alpha',b'_\nu}(\mathbf{r}(t')) \ .
\end{split}
\label{eq:kernel-ini}
\end{equation}
The polarisation matrix $\Pi_{b_\nu,b'_\nu}(t-t')$ entering the above definitions is obtained
from the eigenstates and eigenvalues of the dynamical matrix of the corresponding
bath $\nu$ as follows:
\begin{equation}
\Pi_{b_\nu,b'_\nu}(t-t')
=
\sum_{\lambda}\frac{v^{(\lambda)}_{b_\nu} v^{(\lambda)}_{b'_\nu}}{\omega_{\nu,\lambda}^{2}}
\cos\left(\omega_{\nu,\lambda}(t-t')\right)
\label{eq:PImatrix_time}
\end{equation}

Finally, the stochastic (and hence non-conservative) forces $\eta^{(\nu)}_{i\alpha}(t;{\bf r})$ 
are functions of the initial positions and velocities of the DOF of the bath $\nu$.
Following Refs.~[\onlinecite{Kantorovich:2008,Stella:2014}],
we can now assume that each bath $\nu$,
described by the combined Lagrangian $\mathcal{L}_{{\rm bath},(\nu)}+\mathcal{L}_{{\rm int},(\nu)}$,
is in thermodynamic equilibrium at temperature $T_\nu$. Therefore, the stochastic
forces $\eta^{(\nu)}_{i\alpha}(t;\mathbf{r})$ can be treated as
random variables. From these assumptions, the dissipative forces are well described
by a multi-dimensional Gaussian stochastic process with correlation
functions \cite{Kantorovich:2008,Stella:2014}:
\begin{align}
&\langle \eta^{(\nu)}_{i\alpha}(t;{\bf r})\rangle =0
\label{eq:noise-av-1}\\
&\langle \eta^{(\nu)}_{i\alpha}(t;{\bf r}) \eta^{(\nu')}_{i'\alpha'}(t';\mathbf{r})\rangle  = \delta_{\nu\nu'} k_{B}T 
K^{(\nu)}_{i\alpha,i'\alpha'}(t,t';\mathbf{r}) \ .
\label{eq:noise_correl}
\end{align}

\subsection{Extended Langevin Dynamics with auxiliary DOF}
\label{sec:GLEvDOF}

Following Ref~[\onlinecite{Stella:2014}], we transform the GLE given by Eq.~(\ref{eq:gle})
into a more convenient set of Markovian Langevin dynamics 
(with white noise) by introducing a set of auxiliary DOF for each bath 
\cite{Kupferman:2004,Bao:2004,Luczka:2005,Ceriotti:2009,Ferrario:1979,Marchesoni:1983} .

First we proceed with a mapping of the memory kernel of each bath by transforming the 
$\Pi_{b_\nu,b'_\nu}$ matrices as follows \cite{Stella:2014,Ness:2015}
\begin{equation}
\begin{split}
\Pi_{b_\nu,b'_\nu}(t-t') \rightarrow \\
\sum_{k_\nu=1}^{N_\nu^{\rm aDOF}} c_{b_\nu}^{(k_\nu)} c_{b'_\nu}^{(k_\nu)}
e^{-\vert t-t'\vert/\tau_{k_\nu}} \cos(\omega_{k_\nu}\vert t-t'\vert) \ .
\end{split}
\label{eq:GLE_mapping_vDOF}
\end{equation}

We then introduce two sets of auxiliary DOF (aDOF) $\{s_{\nu,1}^{(k_\nu)}(t), s_{\nu,2}^{(k_\nu)}(t)\}$ 
corresponding to each independent bath $\nu=1,2$.
They are associated with
the corresponding mapping coefficients 
$\{\tau_{k_1},\omega_{k_1},c_{b_1}^{(k_1)}\}$ with $k_1=1,2,...,N_1^{\rm aDOF}$
and $\{\tau_{k_2},\omega_{k_2},c_{b_2}^{(k_2)}\}$ with $k_2=1,2,...,N_2^{\rm aDOF}$.

Note that the frequencies $\omega_{k_\nu}$ used in the mapping Eq.~(\ref{eq:GLE_mapping_vDOF})
of the matrix $\Pi_{b_\nu,b'_\nu}$ 
are not directly related to the eigenvalues $\omega^2_{\nu,\lambda}$ of the dynamical matrix
$D^{(\nu)}_{b_\nu,b'_\nu}$, as explained in detail in Ref.~[\onlinecite{Ness:2015}]. 
There would be a one-to-one correspondence only when the number of aDOF, $N_\nu^{\rm aDOF}$,
is exactly equal to the number of eigenvalues.

For a memory kernel of
the type given in Eq.~(\ref{eq:GLE_mapping_vDOF}), solving the GLE Eq.~(\ref{eq:gle})
is equivalent to solving the following extended variable dynamics \cite{Stella:2014}:
\begin{widetext}
\begin{equation}
\begin{split}
\dot{r}_{i\alpha} & = p_{i\alpha} / m_i \\
\dot{p}_{i\alpha} & =-\frac{\partial\bar{V}}{\partial r_{i\alpha}}
+\sum_{b_1,k_1}\sqrt{\frac{\mu_{l_1}}{\bar\mu_1}} g_{i\alpha,b_1}\left(\mathbf{r}\right)c_{b_1}^{(k_1)} s_{1,1}^{(k_1)}(t)
+\sum_{b_2,k_2}\sqrt{\frac{\mu_{l_2}}{\bar\mu_2}} g_{i\alpha,b_2}\left(\mathbf{r}\right)c_{b_2}^{(k_2)} s_{2,1}^{(k_2)}(t) \\
\dot{s}_{\nu,1}^{(k_\nu)} & =-\frac{s_{\nu,1}^{(k_\nu)}}{\tau_{k_\nu}}
	+\omega_{k_\nu} s_{\nu,2}^{(k_\nu)} - \sum_{i\alpha,b_\nu} \sqrt{\mu_{l_\nu}\bar\mu_\nu}\
	 g_{i\alpha,b_\nu}\left(\mathbf{r}\right) c_{b_\nu}^{(k_\nu)} \dot{r}_{i\alpha}
	+\sqrt{\frac{2k_{B}T_\nu \bar\mu_\nu}{\tau_{k_\nu}}} \xi_{\nu,1}^{(k_\nu)}(t) \ \ \text{for $\nu=1,2$}\\
\dot{s}_{\nu,2}^{(k_\nu)} & =-\frac{s_{\nu,2}^{(k_\nu)}}{\tau_{k_\nu}}-\omega_{k_\nu} s_{\nu,1}^{(k_\nu)}
	+\sqrt{\frac{2k_{B}T_\nu\bar\mu_\nu}{\tau_{k_\nu}}} \xi_{\nu,2}^{(k_\nu)}(t) \ \ \text{for $\nu=1,2$}
\end{split}
\label{eq:extended_GLE}
\end{equation}
\end{widetext}

The corresponding total vector state 
$\bm{X}=\bm{X}[r_{i\alpha},p_{i\alpha},{s}_{1,1}^{(k_1)},{s}_{1,2}^{(k_1)},{s}_{2,1}^{(k_2)},{s}_{2,2}^{(k_2)}]$ 
follows a multivariate Markovian process, where $\xi_{\nu,x}^{(k_\nu)}$ 
are independent Wiener stochastic processes with (white noise) correlation functions
\begin{equation}
\begin{split}
\langle \xi_{\nu,x}^{(k_\nu)}(t) \rangle = 0 \\
\langle \xi_{\nu,x}^{(k_\nu)}(t) \xi_{\nu',x'}^{(k'_{\nu'})}(t') \rangle = 
\delta_{\nu \nu'} \delta_{x x'} \delta_{k_\nu k'_\nu} \delta(t-t') \ .
\end{split}
\label{eq:white_noise}
\end{equation}

We recall that even if the total vector state $\bm{X}$ corresponds to Markovian processes, 
a subset of its components, for example the vector $\bm{\bar X}=\bm{\bar X}[r_{i\alpha},p_{i\alpha}]$, 
is not necessarily following Markovian processes \cite{Gillespie:1996b}.
This was clearly shown in the previous section where the random noise of the corresponding GLE
is actually a coloured noise given by the memory kernel of the GLE.
For such classes of non-Markovian processes (that are components, or functions of one or more components, 
of multivariate Markovian processes) self-consistency
is guaranteed as the Chapman-Kolmogorov equation is satisfied \cite{Gillespie:1996b}.

\subsection{Integration algorithm from a Fokker-Planck approach}
\label{sec:GLE_FP_andco}

Following the scheme given in Ref.~[\onlinecite{Stella:2014}], 
we now develop a Fokker-Planck (FP) approach to derive a set of equations which are
equivalent to the equations of the extended Langevin dynamics given by Eq.~(\ref{eq:extended_GLE}).
We consider the probability distribution function (PDF) 
$P(r_{i\alpha},p_{i\alpha},\{{s}_{\nu,1}^{(k_\nu)},{s}_{\nu,2}^{(k_\nu)}\},t)$
of the total vector state  $\bm{X}$.
Such a PDF follows a FP dynamical equation which can be written as follows
\begin{equation}
\begin{split}
 \dot{P}(r_{i\alpha},p_{i\alpha},\{{s}_{\nu,1}^{(k_\nu)},{s}_{\nu,2}^{(k_\nu)}\},& t) = \\
 -\hat{\mathfrak{L}}_{\rm FP} & P(r_{i\alpha},p_{i\alpha},\{{s}_{\nu,1}^{(k_\nu)},{s}_{\nu,2}^{(k_\nu)}\},t) 
\end{split}
\label{eq:EOM_PDF}
\end{equation}
where $\hat{\mathfrak{L}}_{\rm FP}$ is the corresponding FP Liouvillian.

We split the Liouvillian $\hat{\mathfrak{L}}_{\rm FP}$ in two parts $-$ a conservative 
part  $\hat{\mathfrak{L}}_{\rm cons}$ 
and a dissipative part $\hat{\mathfrak{L}}_{\rm diss}$.
The dynamics generated by the conservative part $\hat{\mathfrak{L}}_{\rm cons}$ corresponds to the EOM
of the DOF and aDOF given in Eq.~(\ref{eq:extended_GLE}) if one omits all the terms containing
the $\tau_{k_\nu}$ parameters \cite{Stella:2014}.
The remaining dissipative part $\hat{\mathfrak{L}}_{\rm diss}$ generates the EOM of the aDOF, given by
the following generic form:
\begin{equation}
\begin{split}
\dot{s}_{\nu,x}^{(k_\nu)}(t) = -\frac{s_{\nu,x}^{(k_\nu)}(t)}{\tau_{k_\nu}}
		+\sqrt{\frac{2k_{B}T_\nu\bar\mu_\nu}{\tau_{k_\nu}}} \xi_{\nu,x}^{(k_\nu)}(t) \ .
\end{split}
\label{eq:EOM_PDF_diss}
\end{equation}
For such a stochastic dynamics there exists an exact integration algorithm 
\cite{Stella:2014,Gillespie:1996}.

In order to obtain an integration algorithm (see details in Appendix \ref{app:algo}), we consider
the time evolution of the PDF over an elementary time-step $\Delta t$
\begin{equation*}
P(\{\dots\},t+\Delta t)=e^{-\hat{\mathfrak{L}}_{\rm FP}\Delta t}P(\{\dots\},t) \ .
\end{equation*}
We use the splitting of $\hat{\mathfrak{L}}_{\rm FP}$ and a second-order Trotter expansion to decompose
the time evolution operator as follows \cite{Note:trotter_decomp}:
\begin{equation*}
e^{-\hat{\mathfrak{L}}_{\rm FP}\Delta t} \sim e^{-\hat{\mathfrak{L}}_{\rm diss}\frac{\Delta t}{2}}
e^{-\hat{\mathfrak{L}}_{\rm cons}\Delta t}e^{-\hat{\mathfrak{L}}_{\rm diss}\frac{\Delta t}{2}} \ .
\label{eq:trotter}
\end{equation*}
The first and last time evolution operators $e^{-\hat{\mathfrak{L}}_{\rm diss}\frac{\Delta t}{2}}$ with
half a time step ${\Delta t}/{2}$ generate steps (A) and (H) in the algorithm given in Appendix \ref{app:algo}.

Furthermore, we use a second Trotter expansion of the term $e^{-\hat{\mathfrak{L}}_{\rm cons}\Delta t}$
by splitting $\hat{\mathfrak{L}}_{\rm cons}$ in two parts $-$ $\hat{\mathfrak{L}}_{r,s_1}$ and
$\hat{\mathfrak{L}}_{p,s_2}$. 
The part in $\hat{\mathfrak{L}}_{p,s_2}$ generates the time evolution of the 
system DOF $p_{i\alpha}$ and of the aDOF $s_{\nu,2}^{(k_\nu)}$ over half a timestep
${\Delta t}/{2}$, see steps (C) and step (G) in Appendix \ref{app:algo}. 
The part in $\hat{\mathfrak{L}}_{r,s_1}$ generates the time evolution 
of the system DOF $r_{i\alpha}$ and of the aDOF $s_{\nu,1}^{(k_\nu)}$ over ${\Delta t}$, 
see steps (D) and (F) in Appendix \ref{app:algo}.

\subsection{Calculation of the polarisation matrices $\Pi_{b_\nu b_\nu^\prime}$ }
\label{sec:calc_PI_mat}

In order to perform the mapping given by Eq.~(\ref{eq:GLE_mapping_vDOF}), we first
Fourier transform the equation into:
\begin{equation}
\begin{split}
\Pi_{b_\nu,b_\nu^\prime}(\omega)=
\sum_{k_\nu} c_{b_\nu}^{(k_\nu)}c_{b_\nu^\prime}^{(k_\nu)}
\left[\frac{\tau_{k_\nu}}{1+(\omega-\omega_{k_\nu})^{2}\tau_{k_\nu}^{2}}\right.\\
\left. +\frac{\tau_{k_\nu}}{1+(\omega+\omega_{k_\nu})^{2}\tau_{k_\nu}^{2}}\right] .
\end{split}
\label{eq:mapping_PI_matrix}
\end{equation}

For each bath, the set of parameters $\{c_{b_\nu}^{(k_\nu)},\tau_{k_\nu},\omega_{k_\nu}\}$
is obtained from a fitting procedure (described in detail in Ref.~[\onlinecite{Ness:2015}])
based on the vibrational properties of the bath.
As shown in Ref.~[\onlinecite{Ness:2015}], the polarisation matrices $\Pi_{b_\nu b_\nu^\prime}(\omega)$
are related to the imaginary part of the phonon bath propagator 
$\mathcal{D}_{b_\nu b_\nu^\prime}(\omega)$ as follows:
\begin{equation}
\Pi_{b,b^\prime}(\omega) = - {2}{\rm Im} \mathcal{D}_{b,b^\prime}(\omega) / {\vert\omega\vert} . \nonumber
\end{equation}
The bath propagator is defined from the dynamical matrix $\bm{D}^{(\nu)}$
of the bath $\nu$ 
as
\begin{equation}
\mathcal{D}_{b_\nu,b_\nu^\prime}(\omega) = \left[ \omega^2 \bm{1} - \bm{D}^{(\nu)} + i\varepsilon \right]^{-1}_{b_\nu,b_\nu^\prime} , \nonumber
\end{equation}
where $\varepsilon\rightarrow 0^+$.
We use a real-space method, based on the Lanczos algorithm, to calculate the inverse of the matrix defining
$\mathcal{D}_{b_\nu,b_\nu^\prime}$ and the fitting procedure, described in Ref.~[\onlinecite{Ness:2015}],
to get the values of the parameters $\{c_{b_\nu}^{(k_\nu)},\tau_{k_\nu},\omega_{k_\nu}\}$.

Once the system is defined, the calculations of the dynamical matrices and the mapping procedures are
performed individually for each bath $\nu$.

\subsection{Generalisation to $N_{\rm bath}$ independent baths}
\label{sec:genNbath}

It should be noted that the generalisation to the case of the central system
connected to $N_{\rm bath}$ independent baths is straightforward. 
One can expand the GLE-2B by introducing $N_{\rm bath}$
auxiliary sets of DOF $s_{\nu,1}^{(k_\nu)}$ and $s_{\nu,2}^{(k_\nu)}$. 
The EOM of these
aDOF are given in the 3rd and 4th lines of Eq.~(\ref{eq:extended_GLE}).
The EOM for the momenta of the central system will include the contribution of the $N_{\rm bath}$
baths via the set of aDOF. It is defined by the sum over all the bath indices
instead of just a sum over $\nu=1,2$, i.e.
\begin{equation}
\begin{split}
\dot{p}_{i\alpha}  =-\frac{\partial\bar{V}}{\partial r_{i\alpha}}
+\sum_{\nu=1}^{N_{\rm bath}} \sum_{b_\nu,k_\nu}\sqrt{\frac{\mu_{l_\nu}}{\bar\mu_\nu}} g_{i\alpha,b_\nu}
\left(\mathbf{r}\right)c_{b_\nu}^{(k_\nu)} s_{\nu,1}^{(k_\nu)}(t) .
\end{split}
\label{eq:gle_pia_Nbath}
\end{equation}

The mapping and fitting procedures of the polarisations matrices $\Pi_{b_\nu b_\nu^\prime}$ will be performed 
for each individual bath.

\section{Results for the GLE-2B approach}
\label{sec:gle_ex}

\begin{figure}
\begin{centering}
\includegraphics[width=60mm]{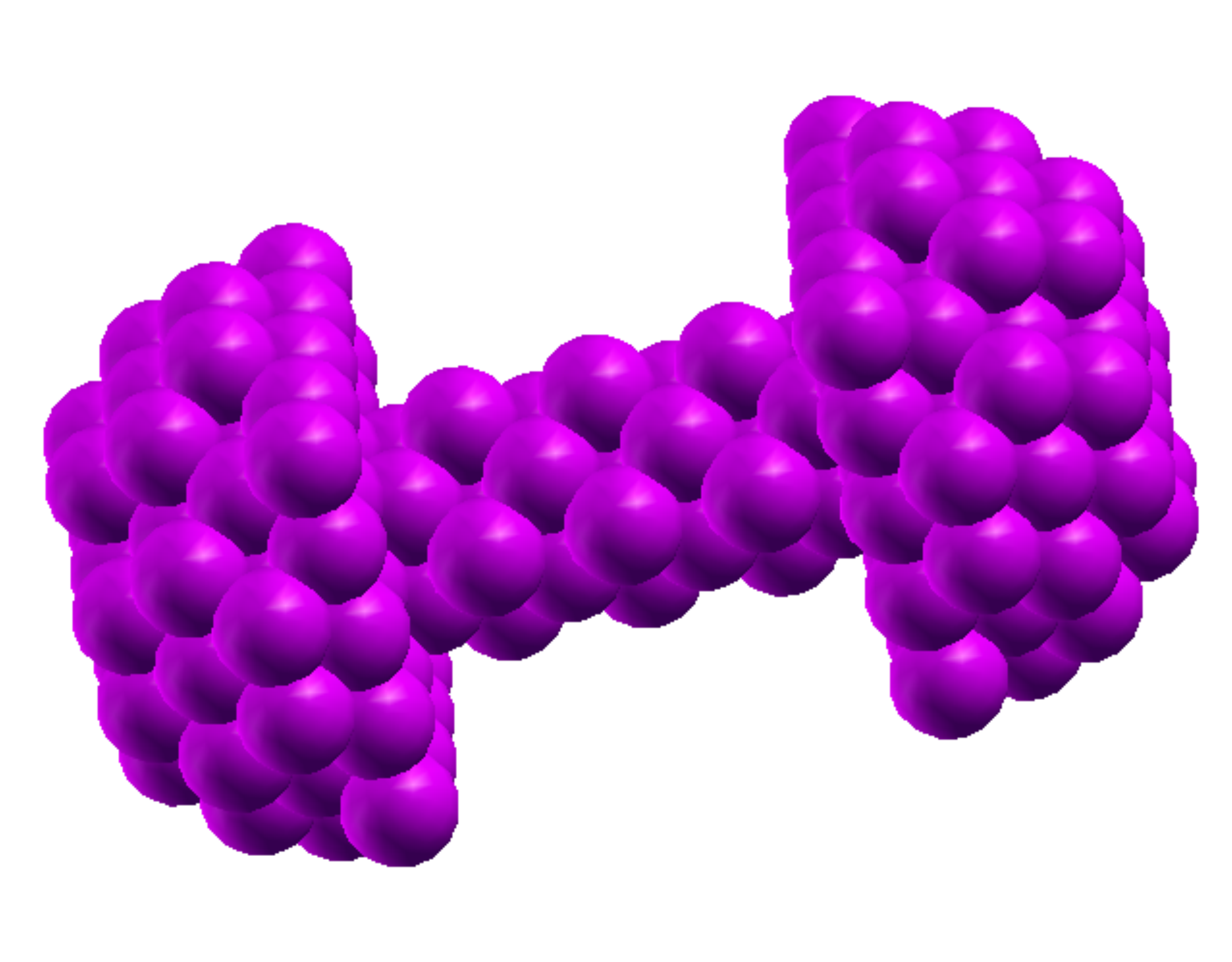}
\end{centering}
\caption{(Colour online)  
Model system for a short cylindric Al wire (central system containing 59 atoms) connected
to two left ($L$) and right ($R$) baths. For the GLE calculations, the $L$ and $R$ bath 
reduced region contains 68 atoms each. 
For the calculations of the dynamical matrix and the mapping of the polarisation matrix, 
larger baths (each containing 203 atoms) were considered \cite{Ness:2015}.
The embedded atom method (EAM) is used for the inter-atomic effective potential.  
For convenience, we now call the baths $\nu=L,R$ instead of $\nu=1,2$.
}
\label{fig:system}
\end{figure}

We now apply the GLE-2B approach to a model system, which consists of a
short Al nanowire connected to two independent baths (represented by two half spheres)
as shown in Figure~\ref{fig:system}. For obvious reasons, we now call the left and
right baths $\nu=L,R$ (instead of $\nu=1,2$).
The electronic transport properties of similar Al nanowires have been studied some
decades ago \cite{Wan:1997,Taraschi:1998,Taylor:2001}, it is now interesting to 
study their thermal transport properties under the proper non-equilibrium conditions.

\begin{figure}
\begin{centering}
\includegraphics[width=70mm]{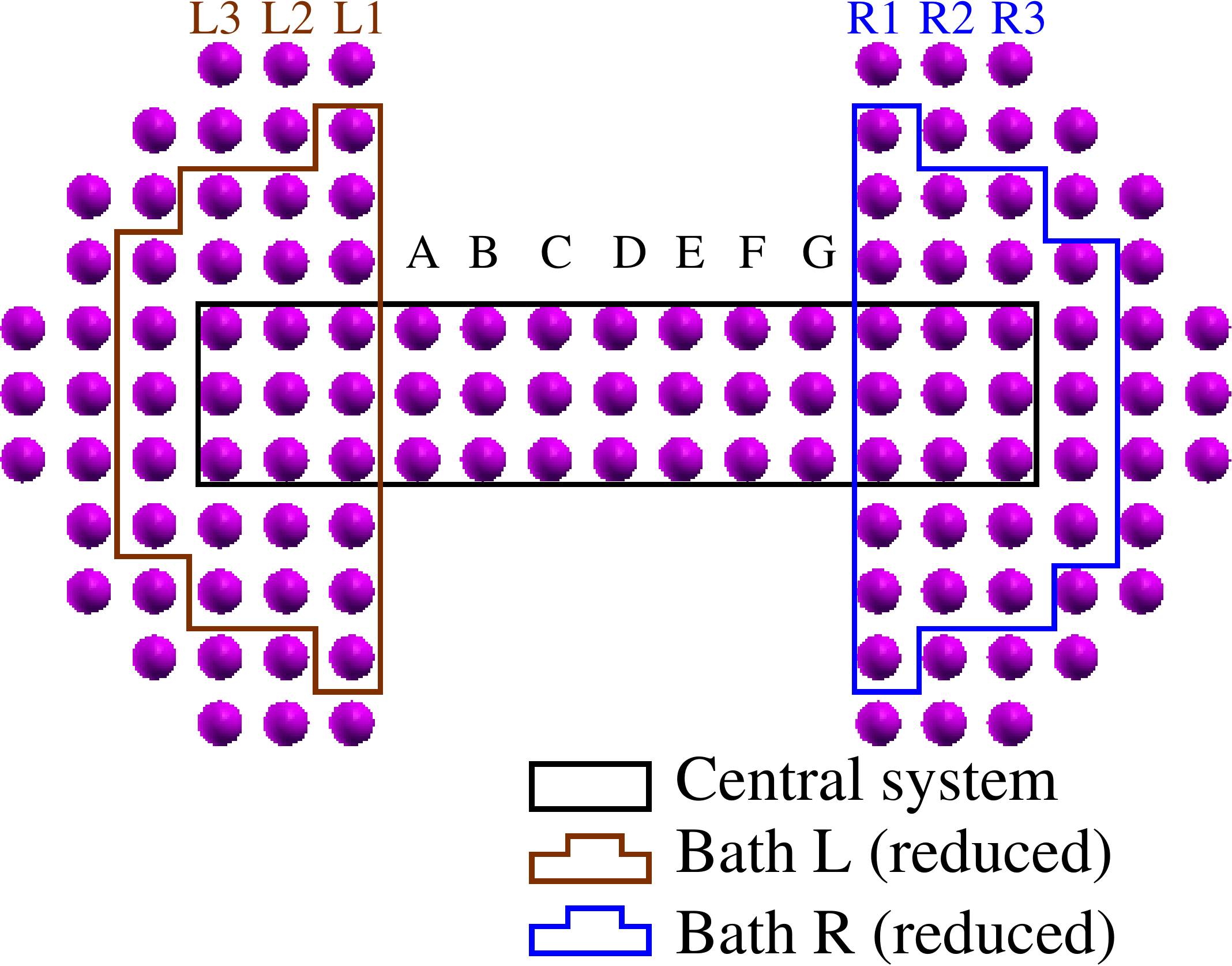}
\end{centering}
\caption{(Colour online)  
Schematic representation of systems under consideration for the GLE calculations. 
The central system consists of 7 layers of Al (labelled A to G) and 3 extra layers
embedded in the left bath (labelled L1, L2 and L3) and 3 extra layers embedded 
in the right bath (labelled R1, R2 and R3).
The layer L3, L1, B, D, F, R1 and R3 (L2, A, C, E, G, R2) contains each 5 (4) Al atoms. 
The bath reduced regions are represented by the brown and blue regions for the left
and right baths respectively. Each of these regions contains 68 atoms. We recall
that, during the GLE simulations, the positions of the bath atoms are fixed. 
The left and right baths are
at their own temperature $T_L$ and $T_R$ respectively.
For the other thermostatting approaches, the bath reduced regions are described by 
their own stochastic dynamics (Langevin-Gauss (LG) 
or Nose-Hoover (NH) thermostatting) at their own temperatures $T_L$ and $T_R$).
The central system (L1-L3, A-G, R1-R3) evolves
according to a Newtonian (NVE) dynamics.
The remaining outer atoms
are kept at fixed positions to ensure the structural stability of the system
in the LG and NH thermostatting calculations.
}
\label{fig:system_gle}
\end{figure}

Figure~\ref{fig:system_gle}
provides more information about the central system (labelling of different layers
of the system) and the bath reduced regions.
The system is built using a fcc lattice and the distance between layer A and layer C 
(Fig.~\ref{fig:system_gle}) corresponds to the lattice parameter of 4.05 \AA.
The Al nanowire (layers A to G) has the length of 12.15 \AA. 
The central system (layers L3-L1, A-G, R1-R3) for which the GLE-2B
simulations are performed has a length of 20.25 \AA.
We have taken the Embedded Atom Method \cite{Daw:1984} to model the metallic Al system. 
The tabulated interatomic potential is provided by the NIST Interatomic 
Potential Repository Project \cite{NISTweb}.

In order to compare to the results obtained with our GLE-2B approach, we also consider
two different thermostatting approaches for the baths. These are more widely used
and consist of stochastic dynamics for the atoms in the bath regions, while the
central region follows the common deterministic Newtonian dynamics 
(see Fig.~\ref{fig:system_gle} and Appendix~\ref{app:algo}).
We consider two stochastic dynamics, i.e. a simple Langevin (Langevin-Gauss LG) 
dynamics \cite{Toton:2010} and the dynamics generated by a Nose-Hoover (NH) thermostat
which are already implemented in the MD code LAMMPS.

For the simple LG approach, the stochastic dynamics for the atoms in the bath 
regions are given by 
\begin{equation}
\dot{\mathbf{p}}_\nu = - \mathbf{\nabla}_{\mathbf{r}_\nu} {V}(\mathbf{r}) - \gamma \mathbf{p}_\nu 
+ \boldsymbol{\xi}^G_\nu \nonumber
\end{equation}
with the momentum vector $\mathbf{p}_\nu$ of the atoms in the bath $\nu=L,R$ and the white noise
vector $\boldsymbol{\xi}^G_\nu$ (following a Gaussian distribution 
with a width related to the temperature $T_\nu$). 
The single parameter $\gamma$ characterises the friction (damping $\tau_{\rm damp}=1/\gamma$) for 
all atoms in the bath regions \cite{Ness:2015}.

For the NH thermostating approach, each is also characterized by a damping parameter 
 $\tau^{\rm NH}_{\rm damp}$ (in unit of time) in the LAMMPS implementation.

As a first application of our GLE-2B, we concentrate on studying the steady-state properties
of the system. For all stochastic dynamics, we consider some initial conditions (values of
the baths' temperatures $T_L$ and $T_R$ and of the velocities of the atoms in the central
system) and let the system evolve in time until the total kinetic energy reaches a plateau, 
i.e. a constant value up to the corresponding thermal fluctuations.
For the different GLE-2B calculations, the system takes roughly 80 ps to reach a
steady state regime (i.e. $\sim$ 40000 time steps for a $\Delta t$ = 2 fs). For the LG
and NH thermostat calculations, the steady state can be reached in less time-steps
since the characteristic damping time is adjustable by the user.

\subsection{Systems at equilibrium}
\label{sec:equi}

In order to validate our GLE-2B approach and the corresponding algorithm, we first need
to check that when the bath temperatures are equal, $T_L=T_R$, we obtain the correct
results for the energy and/or the velocity distributions as expected from equilibrium
statistical mechanics.

\begin{figure}
\begin{centering}
\includegraphics[width=75mm]{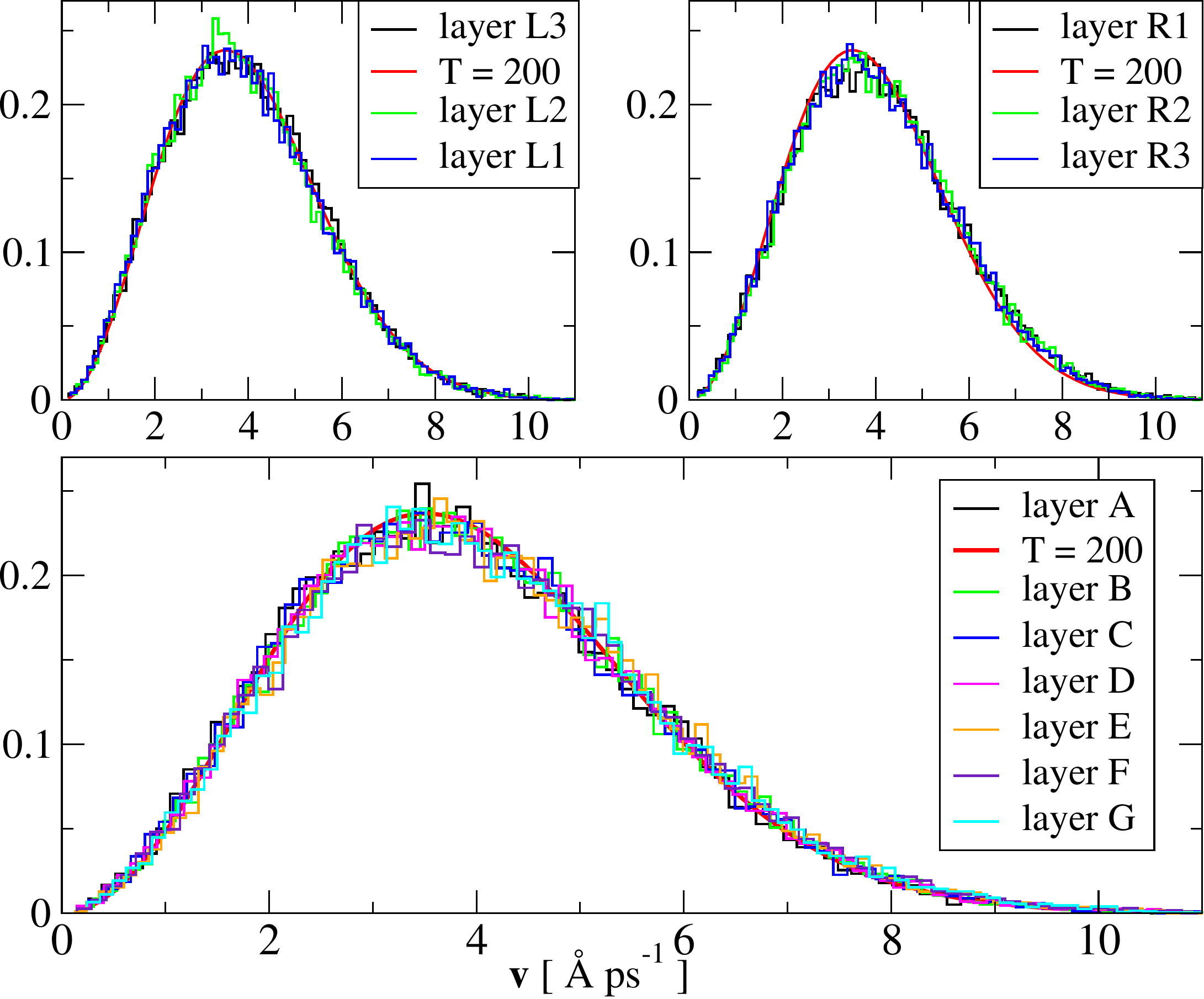}
\end{centering}
\caption{(Colour online)  
Histograms of the velocity distribution for different layers of atoms for the system shown 
in Fig.~\ref{fig:system_gle}.
(Top Left Panel) Layers L3, L2, L1 of the central system.
(Top Right Panel) Layers R1, R2, R3 of the central system.
(Bottom Panel) Layers A to G.
The GLE calculations are performed for $T_L=T_R$ = 200 K.
All distributions (histograms) fit perfectly the equilibrium Maxwell distribution (solid
red line) as expected.
}\label{fig:equi_veloc_gle}
\end{figure}

We have performed different calculations for different temperatures $T=T_L=T_R$ where
$T$ = 100, 150, 200, 300, 400 K. Figure~\ref{fig:equi_veloc_gle} shows the velocity
distribution functions for different atomic layers of the central system \cite{Note:2}.
The results for the velocity distributions show that the system indeed reaches the 
expected thermal equilibrium, where the velocity distributions follow the Maxwell
equilibrium distribution.

It should be noted that for temperatures above $\sim$ 400 K, our model system
shows some structural instabilities \cite{Note:melting}.
Therefore calculations are performed only with temperatures
lower than 400 K.

\begin{figure}
\begin{centering}
\includegraphics[width=70mm]{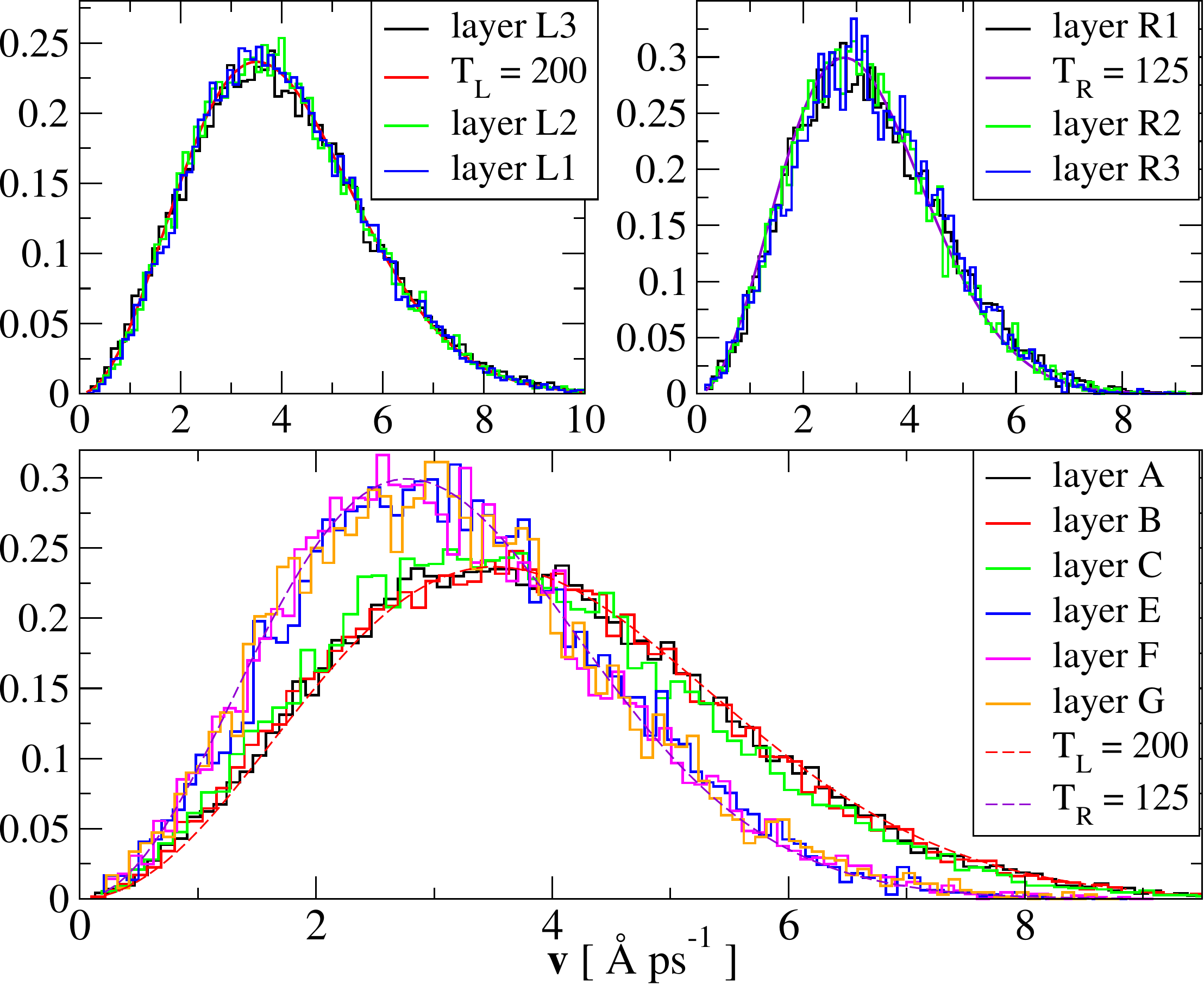}
\end{centering}
\caption{(Colour online)  
Histograms of the velocity distribution for the different layers of atoms in 
the system shown in Fig.~\ref{fig:system_gle}.
The GLE calculations are performed for two different bath temperatures $T_L$ = 200 K and 
$T_R$ = 125 K [\onlinecite{Note:frozenmiddle}].
The central layer D is frozen, i.e. atomic positions fixed, and acts as a perfect thermal barrier.
The distributions (histograms) for the atomic layers on both sides of the frozen layer D, namely 
layers (L3, L2, L1, A, B, C) and layers (E, F, G, R1, R2, R3), fit perfectly 
the equilibrium Maxwell distributions obtained for the two temperatures $T_L$ (lines in 
the top left and bottom panels) and $T_R$ (lines in the top right and bottom panels).
}
\label{fig:veloc_frozenmid_gle}
\end{figure}

To further complement the validity of our approach, we have also performed calculations
for pseudo double equilibrium conditions. 
This is done by considering the two baths at two different temperatures, 
and keeping fixed the atomic coordinates of the central layer D of the central system. 
This creates a thermal decoupling between the two sides, i.e. the frozen layer D acts 
as a perfect reflective
barrier for the thermal transport between the two baths.
Figure~\ref{fig:veloc_frozenmid_gle} shows the results obtained for the velocity distribution
when the bath temperatures are $T_L$ = 200 K and $T_R$ = 125 K [\onlinecite{Note:frozenmiddle}].
One clearly sees that one side of the system (layers L3, L2, L1, A, B, C) has velocity distributions
that are well represented by the equilibrium Maxwell distributions obtained for $T_L$,
while the other side (layers E, F, G, R1, R2, R3) has equilibrium velocity distributions obtained from $T_R$.
Similar results are also obtained when using the LG or NH thermostatting procedures for the bath
stochastic dynamics.  

With this preliminary set of calculations, we feel confident that our approach and 
its numerical implementation in the MD code LAMMPS are correct and, therefore, 
we move onto discussing our out of equilibrium calculations.

\subsection{Non-equilibrium conditions}
\label{sec:NE}

In this section, we consider the proper non-equilibrium conditions when the 
temperatures of the two baths are different $T_L \ne T_R$.
More specifically, we consider the steady state regime when, after some relaxation
time, the total kinetic energy of the system reaches a ``constant'' value (up to the
thermal fluctuations).

In a nanowire connecting two thermal baths when $T_L \ne T_R$, it has been shown
that a temperature gradient may or may not build up across the system.
\HN{Temperature profile measurements are usually performed, by using thermal probe 
AFM, on mesoscopic scale systems (a few microns in length) or
on multiwall carbon nanotubes (14 nm diameter and 4 microns length) \cite{Small:2003}.
Such measurements have not yet been performed on nanoscale objects.
However, a lot of theoretical work on model systems can be found in the literature.
In the following paragraphs, we will review briefly the causes of the presence or absence
of a temperature gradient as reported in previous theoretical studies.}

Since the seminal work of Rieder {\it et al.} [\onlinecite{Rieder:1967}],
it has been known that in an one-dimensional homogeneous harmonic system (also referred to as
an integrable system), the thermal conductivity diverges in the 
thermodynamic limit. No temperature gradient is formed in the bulk of
the system, since the dominating energy ``carriers'' are not scattered 
and propagate ballistically.
A large variety of harmonic (integrable) systems have been studied in 
the classical \cite{Rieder:1967,Rich:1975,Spohn:1977,Casati:1979,Hu:2000,Segal:2008,Kannan:2012,Saaskilahti:2012,Landi:2013}
and quantum \cite{Spohn:1977,Zurcher:1990,Saito:1996,Saito:2000,Dahr:2003,Gaul:2007,Asadian:2013}
limits, using analytical and/or numerical approaches.
All these studies show that there is no temperature gradient inside
the system (except small regions in the vicinity of the contacts
between the central system and the baths).
One usually obtains a constant-temperature profile 
\cite{Rieder:1967,Rich:1975,Spohn:1977,Casati:1979,Hu:2000,Kannan:2012,Landi:2013,
Zurcher:1990,Saito:1996,Saito:2000,Dahr:2003,Gaul:2007,Asadian:2013}
in the central system
around the averaged temperature $T_{\rm av}=(T_L + T_R)/2$.
  
On the other hand, in classical or quantum non-integrable 
systems, a temperature gradient is formed inside the system.
The temperature gradient is uniform, and the heat conductivity 
is finite (it means that these systems obey Fourier's law). 
The general trend is that one has to break
the integrability of the system in order to build up a temperature
gradient there.
This can be done by introducing: 
(a) any form of anharmonicity 
\cite{Jackson:1968,Bolsterli:1970,Nakazawa:1970,Spohn:1977,Eckmann:1999,Hatano:1999,Hu:2000,
Zhang:2002,Pereira:2004,Mai:2006,Bricmont:2007,
Gaul:2007,Segal:2008,Hu:2010,Giberti:2011,Pereira:2011,Saaskilahti:2012,Shah:2013,Landi:2013},
(b) 
extra local stochastic processes \cite{Bolsterli:1970,Rich:1975,Spohn:1977,Zurcher:1990,
Pereira:2004,Bernardin:2005,Kannan:2012,Landi:2013}
or extra collision processes \cite{Lepri:2009,Bernardin:2012},
(c) dephasing for quantum systems \cite{Davis:1978,Saito:1996}, or 
(d) mode coupling for classical systems \cite{Wang:2004}.

The introduction of topological/configurational defects \cite{Jackson:1968,Rubin:1971,Tsironis:1999}  
or disorder \cite{Rubin:1971,Rich:1975,Kipnis:1982,Zhang:2002,Kannan:2012}
in harmonic systems can also lead to the build up of a temperature gradient.
This result can  be understood from the fact that defect/disorder introduces
some form of localisation of the vibration modes \cite{Nakazawa:1970,Stoneham:1975}. 
Such modes do not favour ballistic transport, as phonons get scattered by impurities or 
boundaries \cite{Nakazawa:1970}. Hence the system has a finite thermal conductivity 
and presents a temperature gradient across itself.

\begin{figure}
\begin{centering}
\includegraphics[width=70mm]{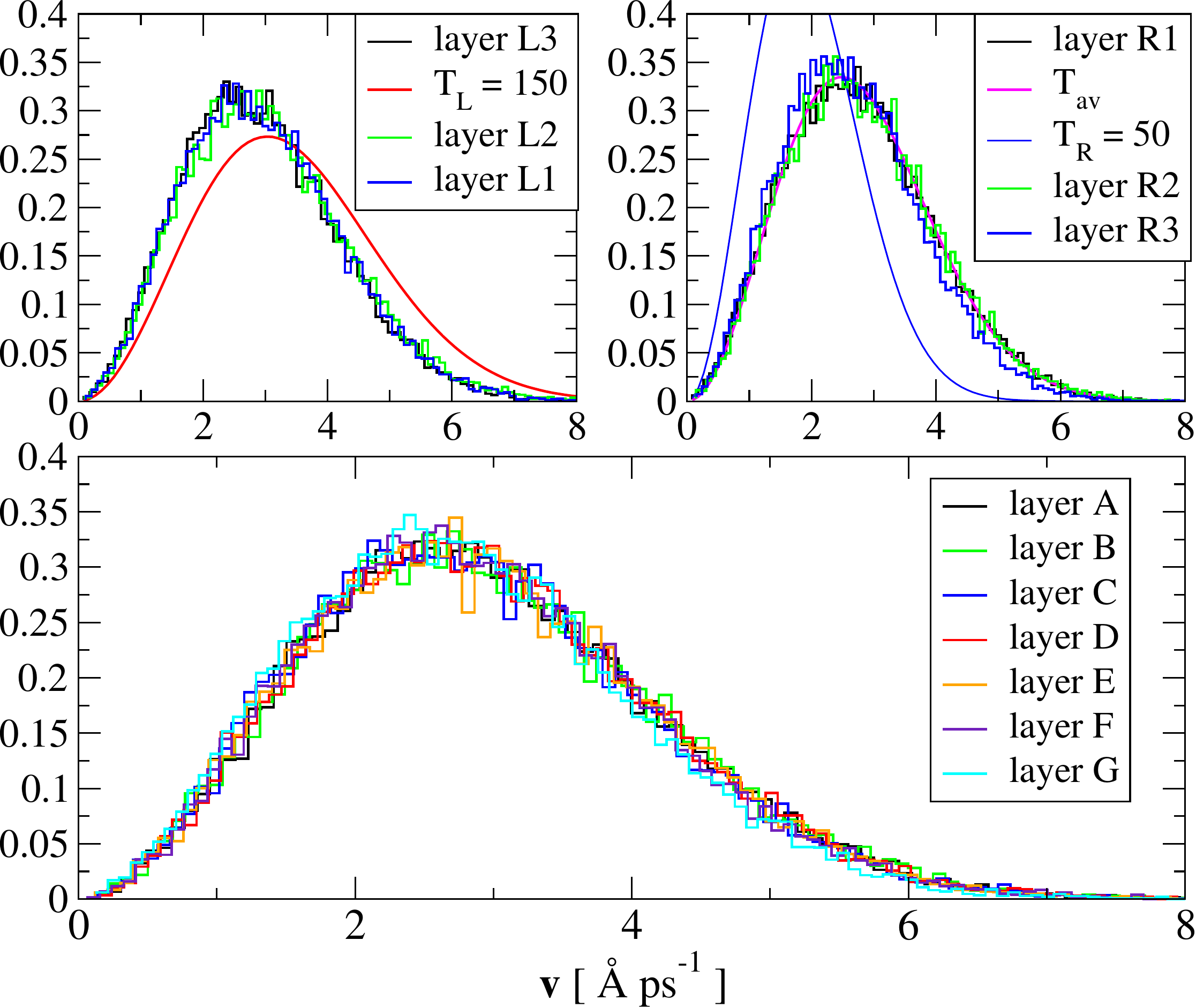}
\end{centering}
\caption{(Colour online)  
Histograms of the velocity distribution for different layers of atoms for the system shown in Fig.~\ref{fig:system_gle}.
The GLE calculations are performed for proper non-equilibrium conditions with the bath temperatures being
$T_L$ = 150 K and $T_R$ = 50 K.
The lineshape of the different distributions (histograms) is similar to that of the
equilibrium Maxwell-Boltzmann distribution. All distributions fit well onto the Maxwell distribution
obtained from the average temperature  $T_{\rm av}=(T_L + T_R)/2=$ 100 K.
}
\label{fig:veloc_ne_gle_TL150TR050}
\end{figure}

We now test our GLE-2B method to investigate the thermal transport properties of the Al nanowire
in the context of ballistic versus diffusive transport regimes.
We consider different temperatures $T_L$ and $T_R$ for the left
and right baths respectively. The temperatures are chosen between
50 and 300 K.

Figure~\ref{fig:veloc_ne_gle_TL150TR050} shows the velocity distribution for a non-equilibrium
calculation performed with $T_L$ = 150 K and $T_R$ = 50 K. This is a typical result, and the
following discussion can be applied to other combinations of bath temperatures $T_L$ and $T_R$
(not shown in the paper).
The corresponding time evolution of the total kinetic energy of the central system and its
statistical distribution is shown in Appendix \ref{app:Ekin_andco}.

First of all, we can see that the lineshape of the different velocity histograms for different
atomic layers is similar to the lineshape of the equilibrium Maxwell distribution. Such
a behaviour permits us to define a local temperature for each different layer.

Second, all velocity histograms appear to follow the same equilibrium Maxwell distribution
corresponding to the average temperature  $T_{\rm av}=(T_L + T_R)/2=$ 100 K (even for the layers L3 
and R3 embedded in the $L$ and $R$ bath respectively).
The overall results clearly indicate that, within such a non-equilibrium regime, we are dealing 
with an integrable system, 
essentially a complex harmonic system which perfectly transmits the heat flux between the
two baths.

To simplify the analysis of our calculations, we now consider only the temperature profiles along the
central system. Such profiles are calculated for each layer by fitting the local velocity histograms
onto the Maxwell distribution. We then extract the corresponding local temperature for each 
layer.

\begin{figure}
\begin{centering}
\includegraphics[width=75mm]{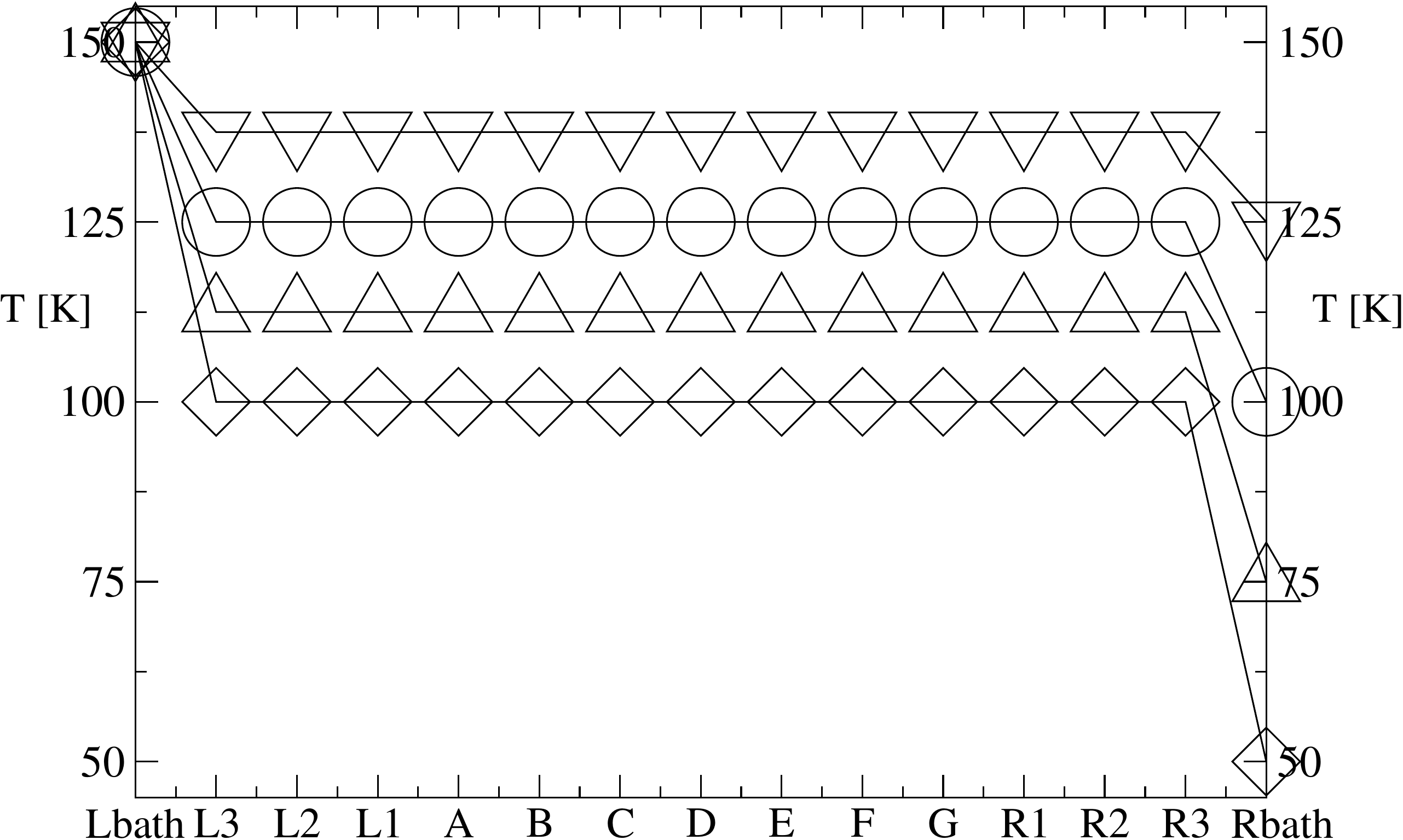}
\end{centering}
\caption{  
Temperature profiles across the central system for different bath temperatures. 
The labels Lbath (Rbath) represent the temperatures 
at the left (right) bath.
The other labels indicate different layers in the central system. For different sets of $T_{L,R}$,
there is no temperature gradient in the system, i.e. it behaves like a perfectly integrable system.
There is an inherent uncertainty in the evaluation of the local temperature from a fit to the 
Maxwell distribution. All temperatures have an error bar of $\pm 5$ K. As a visual guide,
this corresponds to the size of the symbols. 
Note that we also have performed calculations swapping $T_L \leftrightarrow T_R$
and obtained the same flat temperature profiles.
}
\label{fig:Tprofile_GLE_harm}
\end{figure}

Figure~\ref{fig:Tprofile_GLE_harm} shows the temperature profiles in the central system for different sets
of temperatures $T_{L,R}$. From the procedure of fitting the velocity histograms to the Maxwell distribution,
we estimate an uncertainty of approximately $\pm 5$ K on the temperature values.
Even in the presence of such an error, one can see that there is no temperature gradient in the central 
system for the different temperatures considered in Fig.~\ref{fig:Tprofile_GLE_harm}.
This implies that, for such temperatures, the system behaves like a perfect harmonic (integrable) thermal 
conductor.

We can also compare our GLE-2B approach with the other LG and NH thermostatting approaches 
(Fig.~\ref{fig:GLE_and_otherthermostats}). In such approaches, the atoms in the bath
reduced regions are allowed to move and follow a dissipative dynamics ruled by a simple LG
dynamics or by a NH thermostat.
The atoms in the central region (while interacting with themselves and with the moving atoms of 
the two baths) follow a Newtonian NVE dynamics.

\begin{figure}
\begin{centering}
\includegraphics[width=75mm]{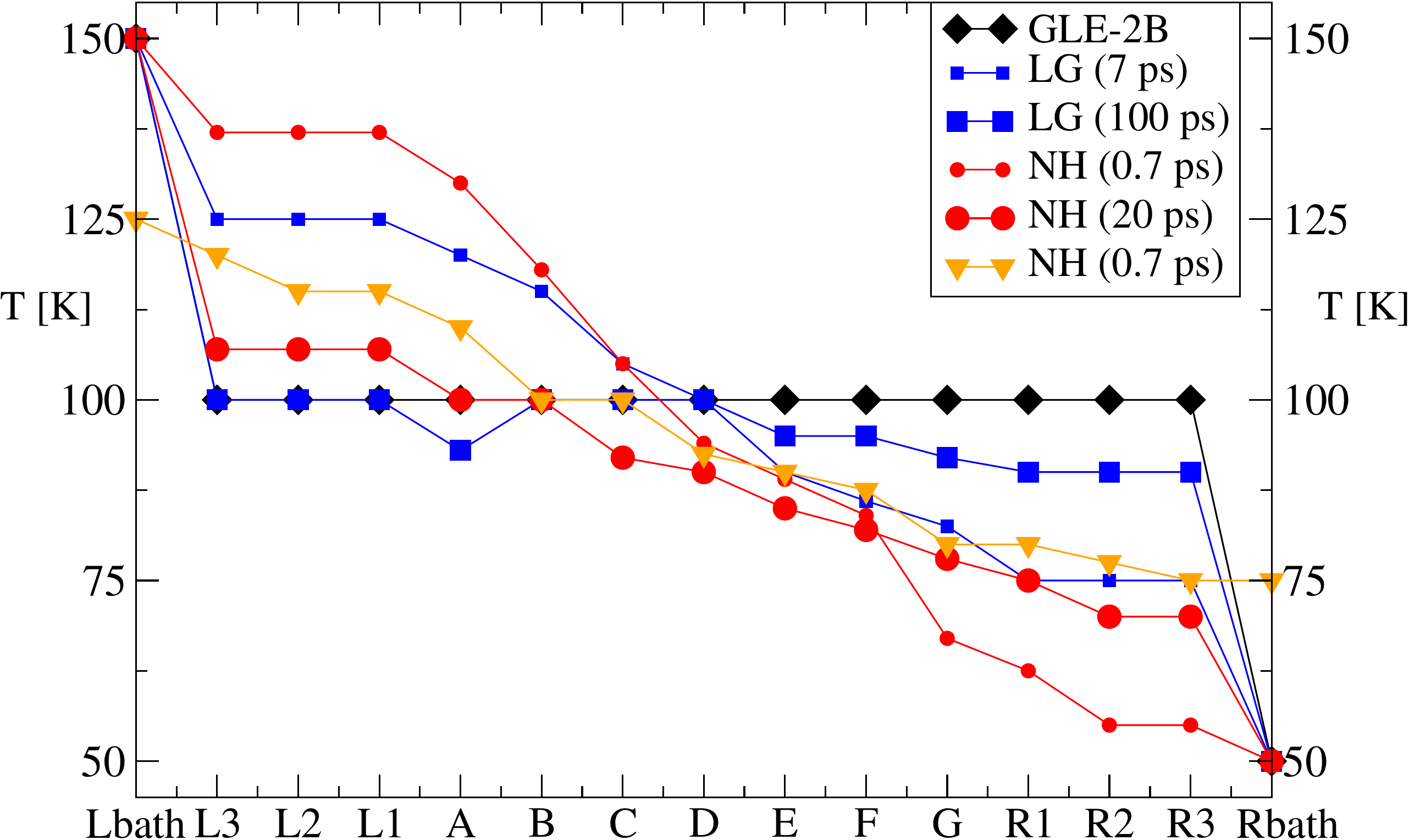}
\end{centering}
\caption{(Colour online)  
Temperature profiles across the system for different bath temperatures $T_L=150$, $T_R=50$. 
Comparison between the GLE-2B approach and the other LG and NH 
thermostatting approaches.
While the GLE-2B calculations provide a uniform temperature profile inside the system, 
the LG and NH thermostats show either the building up of a temperature 
gradient in the central system or a flatter temperature profile depending on
the value used for the damping parameter. 
The LG calculations performed with the damping parameter 
$\tau_{\rm damp}=1/\gamma=7$ ps show a temperature gradient, 
while almost no gradient is obtained for $\tau_{\rm damp}=100$ ps.
The NH calculations show a strong temperature gradient for $\tau^{\rm NH}_{\rm damp}=0.7$ ps,
and a flatter temperature gradient for $\tau^{\rm NH}_{\rm damp}=20$ ps.
There is still a gradient for NH thermostats (with $\tau^{\rm NH}_{\rm damp}=0.7$ ps)
even for smaller $\Delta T$ with  $T_L=125$ and $T_R=75$ (see orange curve).
}
\label{fig:Tprofile_GLE_and_otherthermo}
\end{figure}

\HN{
From the results shown in Figure~\ref{fig:Tprofile_GLE_and_otherthermo}, we can see that
while the  GLE-2B calculations provide a uniform temperature profile inside the system,  
the LG and NH thermostatting approaches display two different kind of behaviour
depending on the chosen value of damping parameter.
Note that in all LG and NH calculations, we have performed the calculations with as many 
time-steps as required to reach a stationary state. The larger the value of the damping
parameter, the longer the dynamics is needed to reach the stationary state
(See also Appendix \ref{app:Ekin_andco} for the time evolution of the kinetic 
energy).

For small values of the damping parameter 
$\tau_{\rm damp}=7$ ps ($\tau^{\rm NH}_{\rm damp}=0.7$ ps), we obtain a temperature
gradient in the central system. The NH thermostats appear to provide an almost perfect 
linear gradient in the  whole system, while the simple Langevin thermostatting corresponds to
a small temperature gradient.

However for larger values of the damping parameter 
$\tau_{\rm damp}=100$ ps ($\tau^{\rm NH}_{\rm damp}=20$ ps), both LG and NH thermostats provide
a  flatter temperature profile. The LG calculations result in a temperature profile almost
similar to that of the GLE-2B calculations.

It is interesting to see that changing the value of the damping can lead to completely
different physical results, i.e. the presence of a temperature gradient or its absence
in the central system.
Such a dilemma does not exist in our GLE-2B approach since it does not contain any
adjustable parameter.

One should note that the small values of the damping parameter has been chosen in
order to reproduce a relaxation of the kinetic energy (for the LG and NH thermostatting)
similar to the relaxation obtained from the GLE-2B approach
(See also Appendix \ref{app:Ekin_andco}).

It seems fair to say that fitting the values of the damping parameter (to reproduce
the evolution of the kinetic energy) is not enough for 
obtaining the same temperature profiles with the LG, NH thermostats and with the GLE-2B approach.

As mentioned above, the presence of a temperature gradient is a signature of the breaking down 
of the integrability of the system, and corresponds to a system with a finite thermal conductance. 
It also implies the introduction of anharmonic effects, configurational disorder or the 
introduction of other
(uncontrolled) random processes.
}

\begin{figure}
\begin{centering}
\includegraphics[width=75mm]{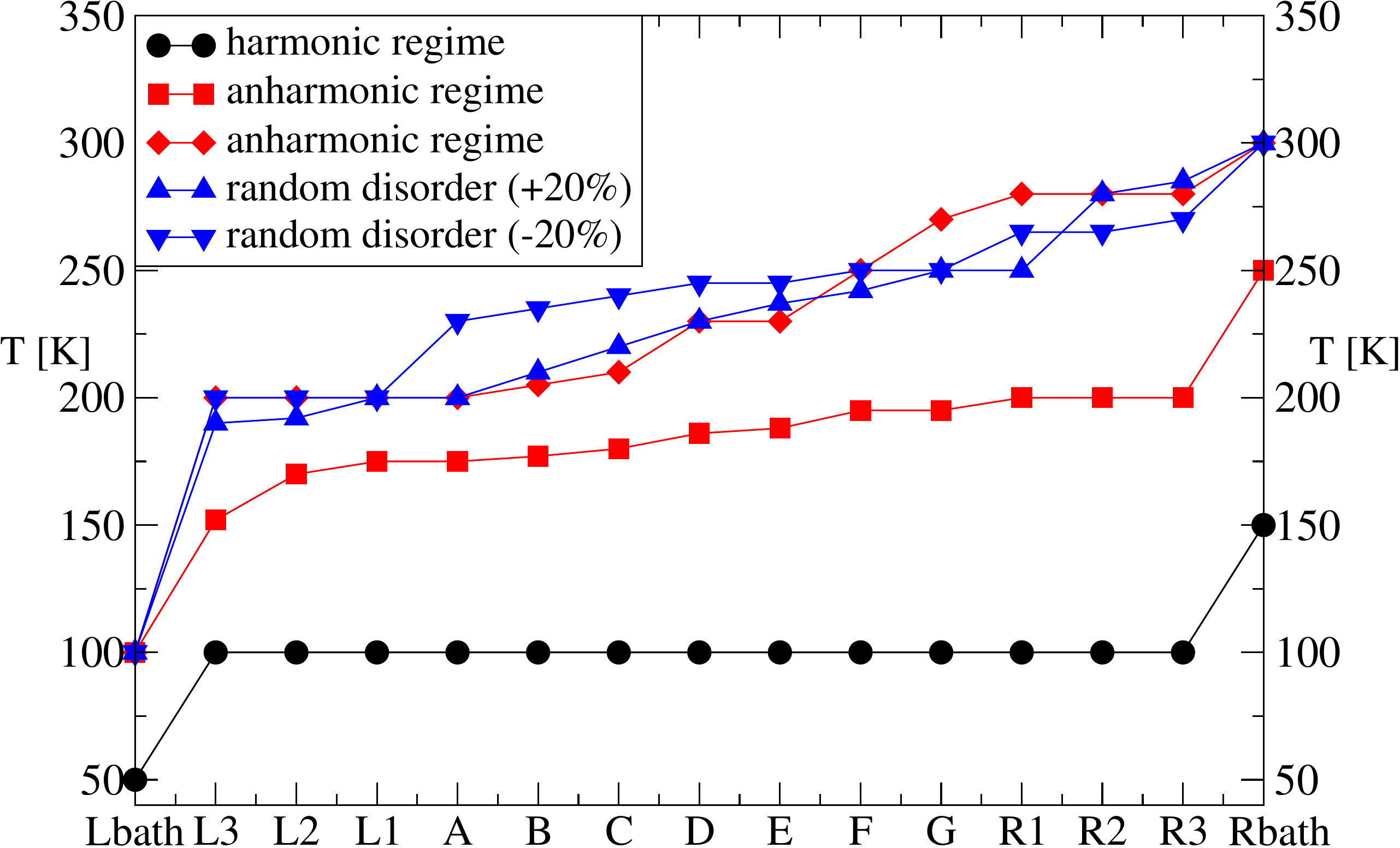}
\end{centering}
\caption{(Colour online)  
Temperature profiles across the system from the GLE-2B calculations
under different $T_{L,R}$ and different atomic masses. It is possible to break the
integrability of the system by considering larger temperatures (red curves) and/or by 
introducing configurational disorder (blue curve). The disorder is introduced by changing the mass
($\pm 20\%$) of some atoms picked up randomly in the central system. For higher 
temperatures, parts of the system are ``driven'' beyond the harmonic limit.
In such cases, the system is not entirely ballistic and presents a finite thermal conductivity, 
leading to the build up of a temperature gradient. 
}
\label{fig:Tprofile_GLE_anharm_desorder}
\end{figure}

In order to understand if one can obtain a temperature gradient with the GLE-2B approach, 
we have
performed calculations in different situations where the harmonicity of the system is
broken. This can be achieved be considering higher temperatures. In these cases, the
atoms of the central system move in an effective potential which goes beyond the harmonic 
potential well and consequently one can obtain a finite temperature gradient, as shown
in Figure~\ref{fig:Tprofile_GLE_anharm_desorder}. One can see that larger deviations 
of the temperatures from the average $T_{\rm av}=(T_L + T_R)/2$ are obtained on the side
of the hotter bath.
One can further increase such effects by introducing artificially some configurational
disorder in the system. 
For that, we have considered the same system and at random we have given one atom in each 
layer L1, A-G and R1 a mass 20$\%$ larger (smaller) than the mass of the other
atoms in the system.
The corresponding temperature profile is shown in Figure~\ref{fig:Tprofile_GLE_anharm_desorder}
and presents, as expected for such a disordered and/or anharmonic system, a finite temperature
gradient.
In Appendix \ref{app:1dmodel} we present calculations for a model of a one-dimensional Al chain.
The GLE-2B calculations obtained for such a simple model confirm, as expected, the analysis
we have performed in this section.

Furthermore, it is also crucial to understand the importance of anharmonic effects.
For that we introduce into the GLE-2B method some form of anharmonic effects in the
baths by taking a finite life-time for the phonon
modes of the bath. 
It means that vibrational excitations do not have an infinite life-time (as
should be the case for an integrable harmonic system) but rather that they decay (dissipate) 
in time. 
The simplest way to treat such effects is to introduce a ``self-energy'' in the
phonon bath propagator
\begin{equation}
\mathcal{D}_{b_\nu,b_\nu^\prime}(\omega) = 
\left[ \omega^2 \bm{1} - \bm{D}^{(\nu)} + \omega\bm{\Sigma}^{\rm anh}(\omega) \right]^{-1}_{b_\nu,b_\nu^\prime} \ .
\nonumber
\end{equation}
Furthermore, we consider that $\bm{\Sigma}^{\rm anh}$ is purely imaginary and
simply modifies the linewidth of the spectral features in  
$\Pi_{b,b^\prime}(\omega) = - {2}{\rm Im} \mathcal{D}_{b,b^\prime}(\omega) / {\vert\omega\vert} $. 
In practice, such effects can be implemented in a rather straightforward way: once the 
mapping of $\Pi_{b,b^\prime}(\omega)$ is established, we take the values of the fitted parameters 
$\tau_{k_\nu}$ and make them smaller. The features of the corresponding vibration modes in
the phonon bath propagator are then broadened \cite{Ness:2015}.

\begin{figure}
\begin{centering}
\includegraphics[width=75mm]{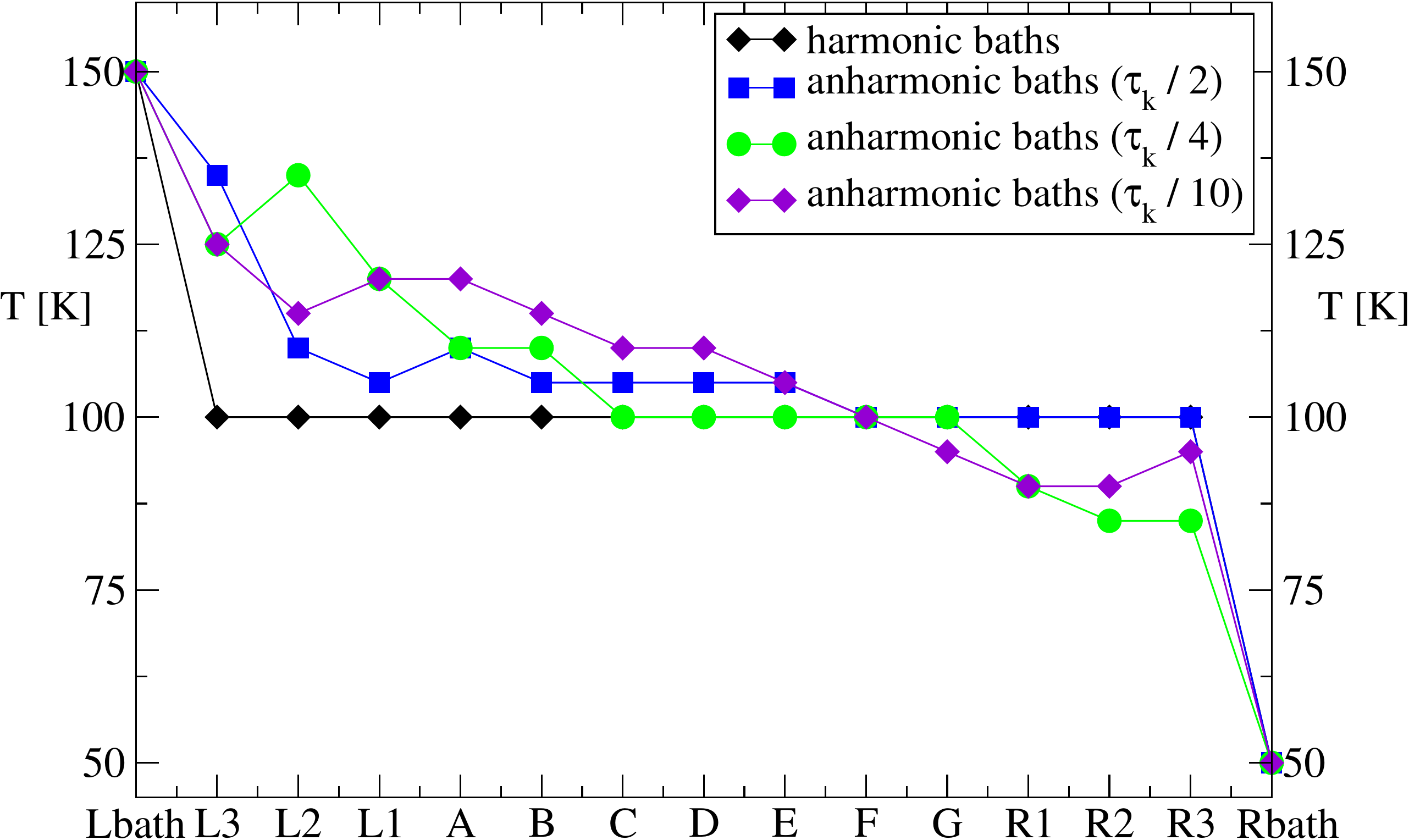}
\end{centering}
\caption{(Colour online)  
Temperature profiles across the system from the GLE-2B calculations. A simplified form
of anharmonic effects in the baths is introduced by modifying the values of the fitted
parameters $\tau_{k_\nu}$. Such anharmonic effects lead to a small temperature gradient
on the side of the hotter bath ($L$ side). 
The building up of the temperature gradient is however not as strong as for the
LG/NH thermostatting calculations.
 }
\label{fig:Tprofile_GLE_anharm_baths}
\end{figure}

The results of such calculations, shown in Figure~\ref{fig:Tprofile_GLE_anharm_baths}, 
suggest that anharmonic effects in the baths tend to favour the build up of the temperature 
gradient across the system.

Finally we would like to add that short nanowires can be ballistic (harmonic regime)
for a range of temperatures $T_{L,R}$. 
However, such a behaviour might not be true for much larger (longer) and strongly 
heterogeneous systems \cite{Meier:2014}. 
Indeed, in such systems, one may expect to observe the presence of disorder, of more localized 
vibrational modes or, more importantly, of vibrational mode ``mixing'' effects
(interaction between phonons) 
which lead to the building up of a temperature gradient in the system.
For example, the process of mode coupling has been studied in model three-dimensional systems \cite{Wang:2004}
and has been shown to lead to the presence of a temperature gradient in the system.

\section{Conclusion}
\label{sec:ccl}

We have developed a Generalised Langevin Equation (GLE) approach to treat 
non-equilibrium conditions when a central classical region is connected to 
two realistic thermal baths at two different temperatures. The method
is called GLE-2B for  Generalised Langevin Equation with two baths.
Following the original GLE approach \cite{Stella:2014,Ness:2015}, the extended 
Langevin dynamics scheme is modified
to take into account two sets of auxiliary degrees of freedom, each of which 
characterises the vibrational properties of the baths. 
These auxiliary degrees of freedom are then used to solve the non-Markovian 
dissipative dynamics of the central
region. 
We have developed the corresponding algorithm for MD simulations and implemented
it within the MD code LAMMPS.

As a first application, we have studied the heat transport properties of a short 
Al nanowire, that connects the left and the right Al baths, 
in the steady-state regime.
We have mostly considered the establishment of a local
temperature profile within the system when the two bath temperatures are different.
Our results are interpreted in terms of the properties of harmonic versus non-harmonic systems,
and the presence or the absence of defects. 
In agreement with earlier studies, we found that 
in a purely harmonic (ballistic) thermal conductor
(with spatially extended normal modes), there 
is no temperature gradient across the central part of the system. 
Whenever the system presents some form of thermal resistance (finite
conductance) due to anharmonic effects, disorder, or extra random processes,
a temperature gradient is present in the system.
Furthermore a concrete example of such effects
in a model of a one-dimension Al chain is provided in Appendix~\ref{app:1dmodel}.
 
\begin{figure}
\begin{centering}
\includegraphics[width=90mm]{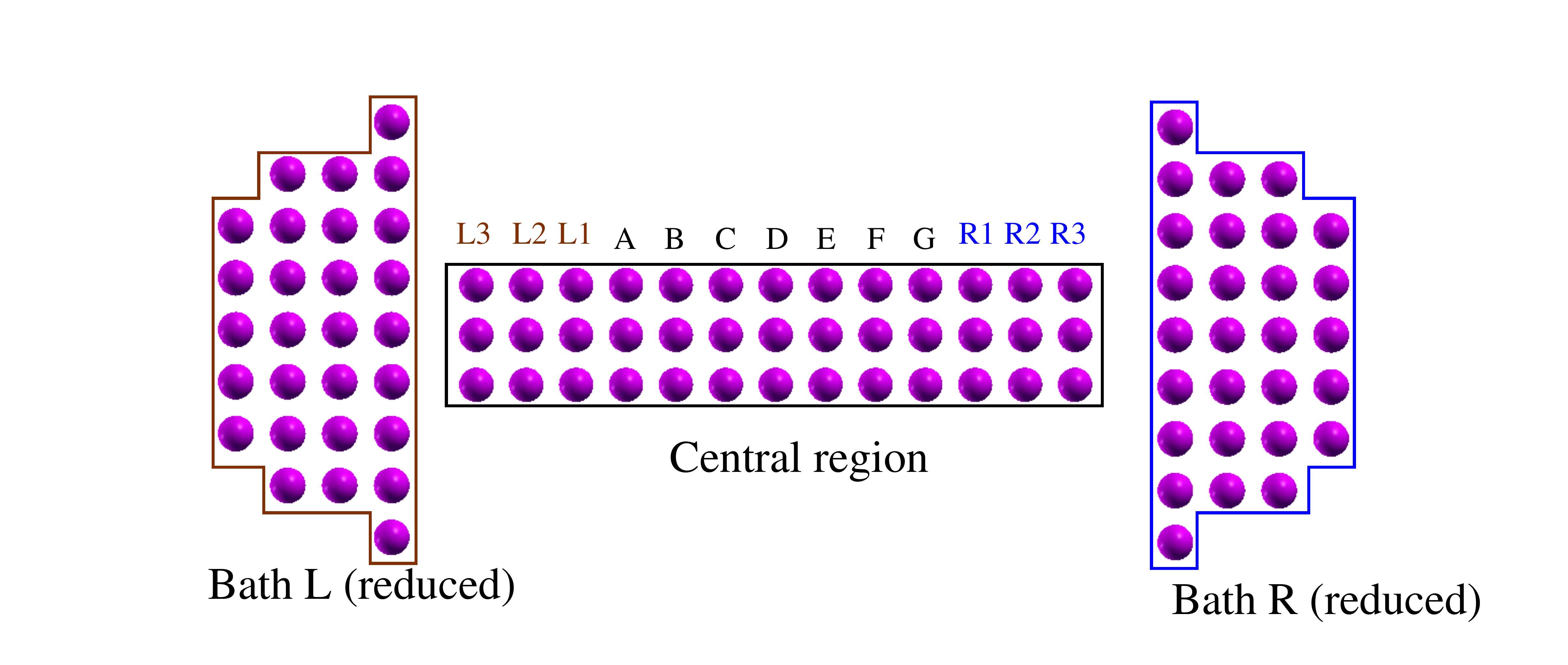}\\
\hspace{8mm}\includegraphics[width=60mm]{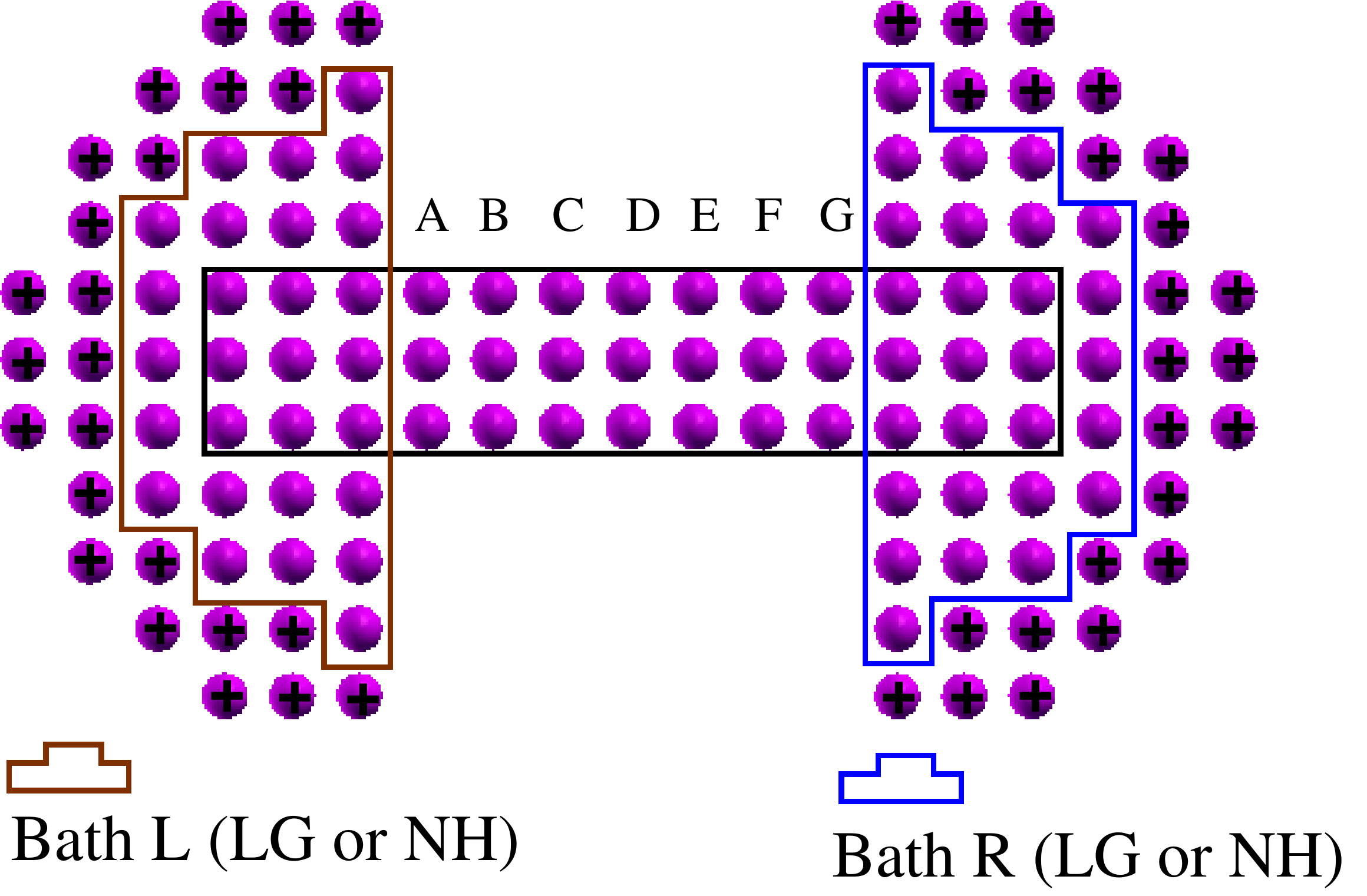}
\end{centering}
\caption{(Colour online)  
Schematic description of the GLE-2B and the LG/NH calculations. 
({\it Upper panel}) the GLE dissipative
dynamics Eq.~(\ref{eq:gle}) is performed only on the atoms of the central region. The
$L$ and $R$ baths (displaced further to the left and right for clarity) enter
into the GLE calculations via the sets of fitting parameters
$\{c_{b_\nu}^{(k_\nu)},\tau_{k_\nu},\omega_{k_\nu}\}$ and 
the $g_{i\alpha,b_\nu}(\{r_{i\alpha}\})$ quantities. In such calculations, the 
positions of the bath atoms are fixed at their equilibrium positions shown in
Fig.~\ref{fig:system_gle}.
({\it Lower panel})
The atoms in the central region follow Newton's EOM and the atoms in the $L$ and
$R$ baths follow a dissipative LG or NH dynamics. The atoms with a black cross
are kept fixed to ensure the overall stability of the system.
The two approaches, GLE-2B and LG/NH, represent two different kinds
of stochastic processes.
}
\label{fig:GLE_and_otherthermostats}
\end{figure}

\HN{
We have also compared the results of the simulations using the GLE-2B approach to the results
of other simulations that were carried out using standard thermostatting 
approaches (based on Markovian Langevin and Nose-Hoover thermostats, see Fig.~\ref{fig:GLE_and_otherthermostats}).
In the latter cases, either a flat temperature profile or a temperature gradient across the 
central system can be obtained depending on the value used for the damping parameter.
Upon the choice of this parameter, two different physical results can be obtained.
Such a dilemma does not exist in the GLE-2B approach as it does not contain
any adjustable parameters.
}

Furthermore, we have shown that the GLE-2B is able to treat, within the same scheme, 
two widely different transport regimes, i.e. systems which have ballistic (with no temperature 
gradient) or 
diffusive (with temperature gradient) thermal transport properties. This is a crucial
point since 
the crossover between ballistic and diffusive transport regimes has been observed 
experimentally \cite{Meier:2014} in organic molecules of different lengths 
connecting two electrodes,
after having been predicted theoretically \cite{Segal:2003}. 

\HN{
Penultimately we would like to add that the GLE-2B has also another advantage over
the more commonly used thermostatting approaches. 
This method has been derived explicitly in order to be able to treat
inherently non-equilibrium properties which cannot be simulated (in principle)
by the NH thermostats. 
Furthermore, we have already shown in
Appendix D of Ref.~[\onlinecite{Stella:2014}] that we can derive
the GLE dynamics with a coloured noise which is not simply proportional 
to the memory kernel (as is the case for the classical limit of the
equilibrium fluctuation-dissipation theorem). This means that quantum
effects of the baths can be incorporated in the GLE dynamics. The importance
of such effects has been considered in Refs.[\onlinecite{Wang:2007},\onlinecite{Ceriotti:2009b}].
Finally, it should be noticed that our GLE-2B approach is also perfectly appropriate to study
time-dependent phenomena. Such interesting phenomena, which involve proper dynamical behaviour 
of systems, will be the subject of future studies.
}

\begin{acknowledgements}

We acknowledge financial support from the UK EPSRC, under Grant No. EP/J019259/1.
HN, CDL and LK acknowledge the stimulating research environment provided by the 
EPSRC Centre for Doctoral Training in Cross-Disciplinary Approaches to 
Non-Equilibrium Systems (CANES, EP/L015854/1).  
Finally, AG would like to acknowledge the Department of Physics at King's College 
London for funding the summer internship which resulted in her contribution 
to this project. 
\HN{
Finally the authors thank one of the Referees for a careful and critical
analysis of our results and for suggestions that strengthened the value of
the present work.
}
\end{acknowledgements}

\appendix

\section{Verlet-type algorithm for the extended Langevin dynamics with two baths}
\label{app:algo}

Following the Markovian equations derived in Section~\ref{sec:GLEvDOF} and 
prescriptions given in Ref.~[\onlinecite{Stella:2014}], we use the following algorithm for a
single time-step $\Delta t$.

The algorithm is derived, in a manner similar to the Verlet algorithm, from a different splitting 
and a Trotter-like decomposition of the 
total Liouvillian for the extended Langevin dynamics of the system DOF, $r_{i\alpha}$, and the auxiliary DOF
$s_{1,2}^{(k_\nu)}$ associated with the two independent baths $\nu=1,2$.
{Such a decomposition has been shown to provide a more appropriate description of the velocity correlation 
functions \cite{Leimkuhler:2013}.}

{Algorithm:}
\begin{equation}
\begin{split}
& \text{(A) Randomise and propagate the aDOF} \\
& s_{\nu,x}^{(k_\nu)} \leftarrow a_{k_\nu} s_{\nu,x}^{(k_\nu)} + b_{k_\nu} \xi_{\nu,x}^{(k_\nu)} \\
& \text{for all $k_\nu$ and $\nu=1,2$ and $x=1,2$.} \\
& \text{(B) Calculate all $g_{i\alpha,b_\nu}(\{r_{i\alpha}\})$ from the} \\
& \text{derivatives of $f_{b_\nu}(\{r_{i\alpha}\})$} \\
& \text{(C) Propagate the DOF and aDOF over $\Delta t / 2$} \\
& v_{i\alpha} \leftarrow v_{i\alpha} + \left( f_{i\alpha} + f_{i\alpha}^{\rm pol} + f_{i\alpha}^{p{\rm GLE}}\right) \frac{\Delta t}{2 m_i}\\
& s_{\nu,2}^{(k_\nu)} \leftarrow s_{\nu,2}^{(k_\nu)}-\omega_{k_\nu} s_{\nu,1}^{(k_\nu)}\frac{\Delta t}{2}\\
& \text{(D) Propagate the DOF over $\Delta t$} \\
& r_{i\alpha} \leftarrow r_{i\alpha}+v_{i\alpha} \Delta t\\
& \text{(E) Recalculate all $g_{i\alpha,b_\nu}(\{r_{i\alpha}\})$ from the} \\
& \text{derivatives of $f_{b_\nu}(\{r_{i\alpha}\})$} \\
& \text{(F) Propagate the aDOF over $\Delta t$} \\
& s_{\nu,1}^{(k_\nu)} \leftarrow s_{\nu,1}^{(k_\nu)}+\left(\omega_{k_\nu} s_{\nu,2}^{(k_\nu)} + f_{k_\nu}^{s{\rm GLE}}\right)\Delta t\\
& \text{(G) Propagate the DOF and aDOF over $\Delta t / 2$} \\
& v_{i\alpha} \leftarrow v_{i\alpha} + \left( f_{i\alpha} + f_{i\alpha}^{\rm pol} + f_{i\alpha}^{p{\rm GLE}}\right) \frac{\Delta t}{2 m_i}\\
& s_{\nu,2}^{(k_\nu)} \leftarrow s_{\nu,2}^{(k_\nu)}-\omega_{k_\nu} s_{\nu,1}^{(k_\nu)}\frac{\Delta t}{2}\\
& \text{(H) Randomise and propagate all the aDOF} \\
& s_{\nu,x}^{(k_\nu)} \leftarrow a_{k_\nu} s_{\nu,x}^{(k_\nu)} + b_{k_\nu} \xi_{\nu,x}^{(k_\nu)} \\
\end{split}
\label{eq:algoC}
\end{equation}
where the different forces, $f_{i\alpha}, f_{i\alpha}^{\rm pol},f_{i\alpha}^{p{\rm GLE}}, f_{k_\nu}^{s{\rm GLE}}$ 
are explained below.
The force 
\begin{equation}
f_{i\alpha} = -\frac{\partial {V}(\mathbf{r})}{\partial r_{i\alpha}}
\label{eq:f}
\end{equation}
is the force acting on the system DOF ${i\alpha}$
due to the interaction between the atoms in the system and in the bath region(s);
the ``polaronic'' force $f_{i\alpha}^{\rm pol}$
\begin{equation}
\begin{split}
f_{i\alpha}^{\rm pol} & = \sum_\nu \sum_{b_\nu,b'_\nu} \sqrt{\mu_{l_\nu} \mu_{l'_\nu}}\ g_{i\alpha,b_\nu}\ 
				\Pi_{b_\nu b'_\nu}(0) f_{b'_\nu} \\
	              & = \sum_\nu \sum_{b_\nu,b'_\nu,k_\nu} \sqrt{\mu_{l_\nu} \mu_{l'_\nu}}\ g_{i\alpha,b_\nu} 
				c_{b_\nu}^{(k_\nu)} c_{b'_\nu}^{(k_\nu)}\ f_{b'_\nu}
\end{split}
\label{eq:fpol}
\end{equation}
(with $b_\nu \equiv l_\nu\gamma$ for DOF the $\nu$-th bath)
is the force acting on the system DOF ${i\alpha}$ due to the interaction between the system and bath regions
which induces a displacement of the positions of the harmonic oscillators characterising the baths.
In Eq.~(\ref{eq:fpol}), we used the fact that $\Pi_{b_\nu b'_\nu}(0)$ is the inverse Fourier transform 
(evaluated at $\tau=0$) of $\Pi_{b_\nu b'_\nu}(\omega)$ given by Eq.~(\ref{eq:mapping_PI_matrix}).

The force $f_{i\alpha}^{p{\rm GLE}}$ acts on the system DOF ${i\alpha}$ and arises from the generalised Langevin
equations:
\begin{equation}
f_{i\alpha}^{p{\rm GLE}} = \sum_\nu \sum_{b_\nu,k_\nu}\sqrt{\frac{\mu_{l_\nu}}{\bar\mu_\nu}}\ 
			   g_{i\alpha,b_\nu}(\{r_{i\alpha}\})\ c_{b_\nu}^{(k_\nu)} s_{1}^{(k_\nu)} \ .
\label{eq:fGLEp}
\end{equation}
The force $f_{k_\nu}^{s{\rm GLE}}$ acts on the aDOF $s_1^{(k_\nu)}$ and also arises from the generalised Langevin
equations
\begin{equation}
f_{k_\nu}^{s{\rm GLE}} = - \sum_{i\alpha,b_\nu} \sqrt{\mu_{l_\nu} \bar\mu_\nu}\ 
			g_{i\alpha,b_\nu}(\{r_{i\alpha}\})\ c_{b_\nu}^{(k_\nu)} v_{i\alpha}
\label{eq:fGLEs}
\end{equation}

The integration of the dissipative part of the dynamics of the aDOF 
(see steps (A) and (F) in the algorithm) 
includes the coefficients $a_{k_\nu}={\rm exp}(-\Delta t / 2 \tau_{k_\nu})$ and 
$b_{k_\nu}=[k_BT_\nu \bar\mu_\nu ( 1 - a_{k_\nu}^2) ]^{1/2}$
and the uncorrelated random variable $\xi_{1,2}^{(k_\nu)}$ corresponding to the white noise.

\HN{
\section{Evolution and statistics of the energy} 
}
\label{app:Ekin_andco}

In this Appendix we show the time evolution of the kinetic energy
and potential energy
for a system under nonequilibrium conditions. We also calculate the
statistical distribution of the kinetic energy
and briefly discuss the convergence
of the numerical calculations for the GLE-2B and LG approaches.

\begin{figure}
\begin{centering}
\includegraphics[width=75mm]{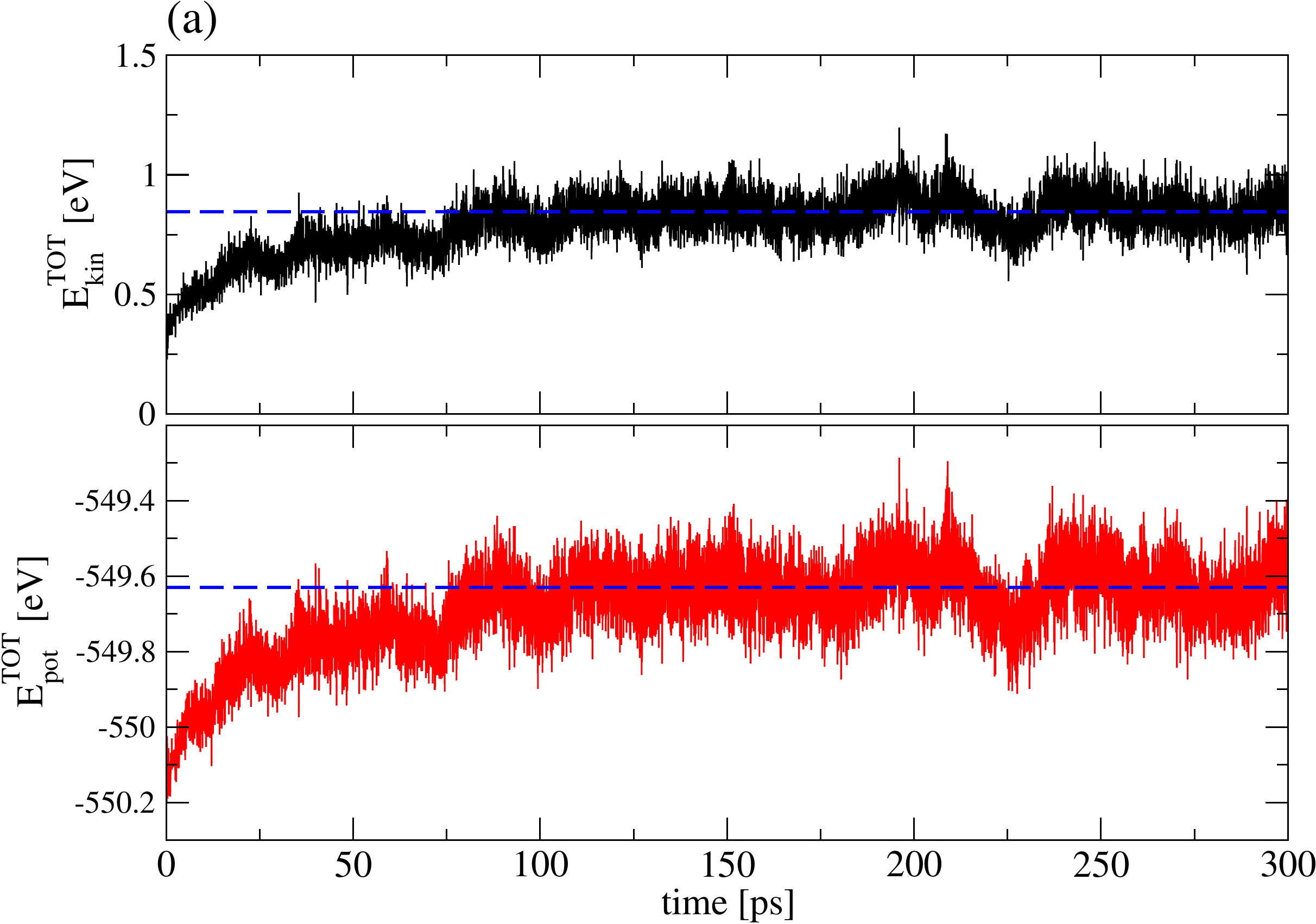}
\includegraphics[width=75mm]{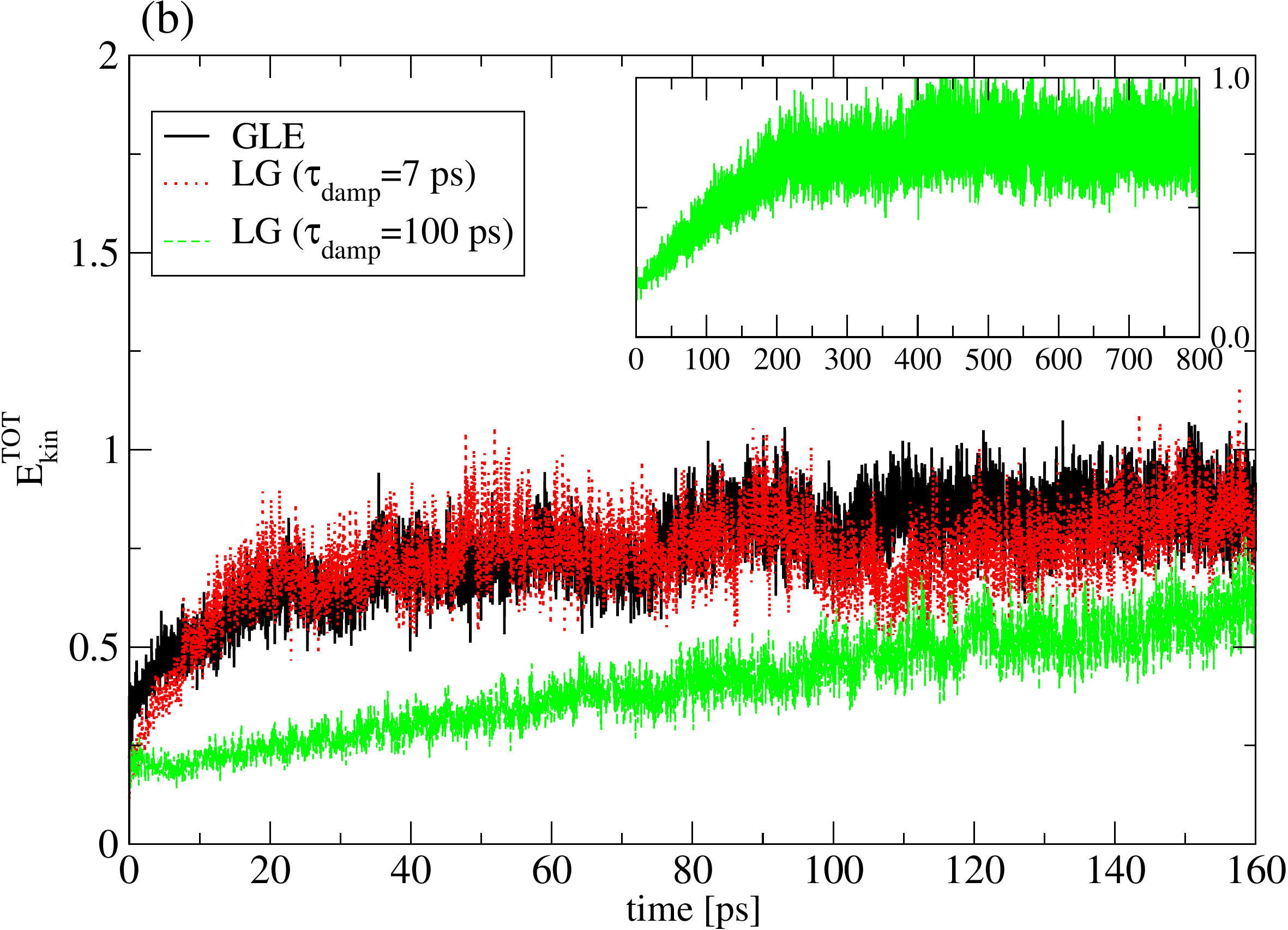}
\end{centering}
\caption{(Colour online) 
(a): Evolution of the total kinetic energy $E^{\rm TOT}_{\rm kin}$ and of the total
potential energy $E^{\rm TOT}_{\rm pot}$ of the system shown in Fig.~\ref{fig:system}
versus time ($\Delta t=2$ fs) [\onlinecite{Note:enerpot}].
The calculations are performed with the bath temperatures $T_L=150$, $T_R=50$. 
The horizontal dashed blue lines correspond to the averages of the energy over the time
range 100 to 300 ps.
The convergence is fairly well obtained after $\sim$ 80 ps (40000 timesteps).
(b): Evolution of the total kinetic energy $E^{\rm TOT}_{\rm kin}$ of the central system 
only (59 atoms from layers L3 to R3).
For the LG calculations, the convergence towards the stationary value occurs after roughly
80 ps for the damping time $\tau_{\rm damp}=7$ ps. 
The stationary state is reached after more 
timesteps, as expected, for the larger damping time $\tau_{\rm damp}=100$ ps (see inset 
where it is apparent that the stationary value is reached after 500 ps or 250000 timesteps).
}
\label{fig:time_evol_EKinTOT}
\end{figure}

Figure~\ref{fig:time_evol_EKinTOT} shows the time evolution 
of the total kinetic energy $E^{\rm TOT}_{\rm kin}$
and of the total potential energy $E^{\rm TOT}_{\rm pot}$
for the non-equilibrium conditions $T_L = 150$ K and $T_R = 50$ K. 
This is a typical result, and the following
discussion can be applied (qualitatively) to other combinations of bath
temperatures [\onlinecite{Note:enerpot}].
For all calculations, $E^{\rm TOT}_{\rm kin}$ reaches an asymptotic 
(nonequilibrium) stationary value.
For the GLE-2B calculations, the stationary regime is obtained after 80 ps
(around 40000 timesteps). While for the LG calculations, the number of timesteps
needed to reach the stationary regime is strongly dependent on the
value used for the damping parameter $\tau_{\rm damp}$.

The asymptotic value of $E^{\rm TOT}_{\rm kin}$ is around 0.8 eV, which
is completely different than the corresponding values expected from
equilibrium and the equipartition principle for the two bath
temperatures, i.e. $\frac{3}{2} Nk_BT_L = 0.38$ eV and $\frac{3}{2} Nk_BT_R = 1.14$ eV.
This is indeed not surprising as the central system is not in an
equilibrium state.

\begin{figure}
\begin{centering}
\includegraphics[width=75mm]{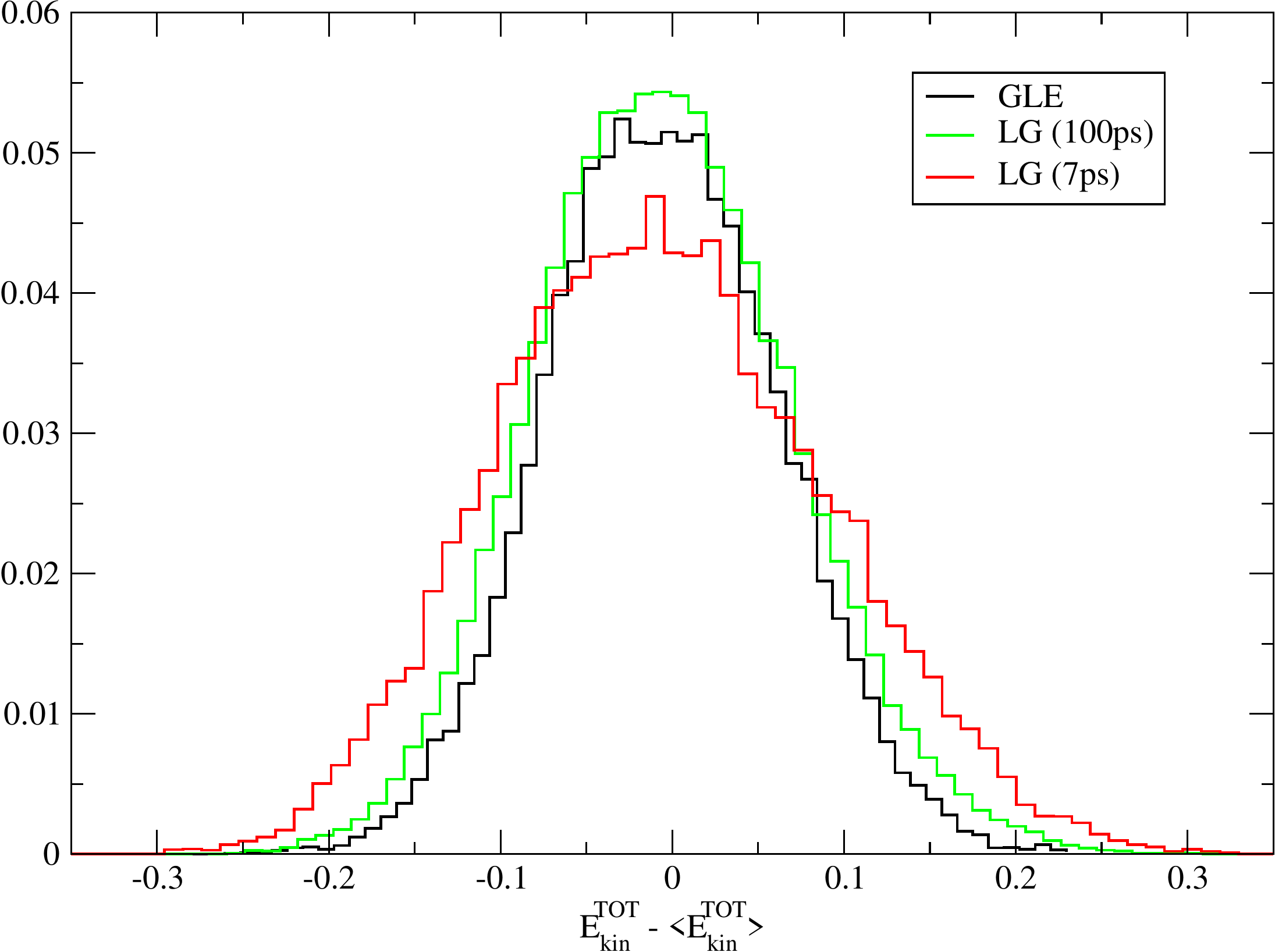}
\end{centering}
\caption{
Histograms of the total kinetic energy of central system containing 59 atoms
for the bath temperatures $T_L=150$, $T_R=50$. The averaged kinetic energy
$\langle E^{\rm TOT}_{\rm kin}\rangle \sim 0.8$ eV is the energy reference.
}
\label{fig:histo_EKinTOT}
\end{figure}

Interestingly, for the nonequilibrium conditions, the choice of the value of the damping parameter 
$\tau_{\rm damp}$ for the LG calculations is crucial to obtain the proper physics. One cannot 
simply use the best $\tau_{\rm damp}$ to reproduce the time 
evolution of the kinetic energy, as done for the equilibrium case \cite{Ness:2015}. 

The influence of the value of $\tau_{\rm damp}$ is reflected in the distribution of the
kinetic energy shown in Figure~\ref{fig:histo_EKinTOT}.
Only the LG calculations performed with a large value of $\tau_{\rm damp}$ reproduce the
distribution of $E^{\rm TOT}_{\rm kin}$ obtained from the GLE-2B approach. The LG
calculations performed with $\tau_{\rm damp}=7$ ps produce a much broader distribution.

Furthermore, in order to obtain the correct temperature profile given by the GLE-2B approach, 
one needs to use a large $\tau_{\rm damp}$ in the LG approach.
The LG calculations made with $\tau_{\rm damp}=7$ ps, which result in a gradient in
the temperature profile (i.e. a completely different physics behaviour, see 
Fig.~\ref{fig:Tprofile_GLE_and_otherthermo}), show a  
time evolution of $E^{\rm TOT}_{\rm kin}$ that is much more similar to that obtained 
from the GLE-2B method.
These results confirm that the choice of the adjustable parameter for the LG method is crucial for
being able to simulate the proper physical behaviour. This is also true for the NH thermostatting
approach (results not shown).


\vspace{1cm}
\section{Influence of the coupling to the baths and the system size}
\label{app:bathcoupling}

In this section, we consider another way of coupling the Al nanowire to the
thermal baths.
We treat a system similar to the system shown in Figure \ref{fig:system}
and Figure \ref{fig:system_gle}
by including into the central system only the 7 layers labelled A to G
(system containing 31 atoms).
The atoms of the layers L1, L2 and L3 (R1, R2 and R3) have now been incorporated 
in the L (R) bath regions themselves.

\begin{figure}
\begin{centering}
\includegraphics[width=75mm]{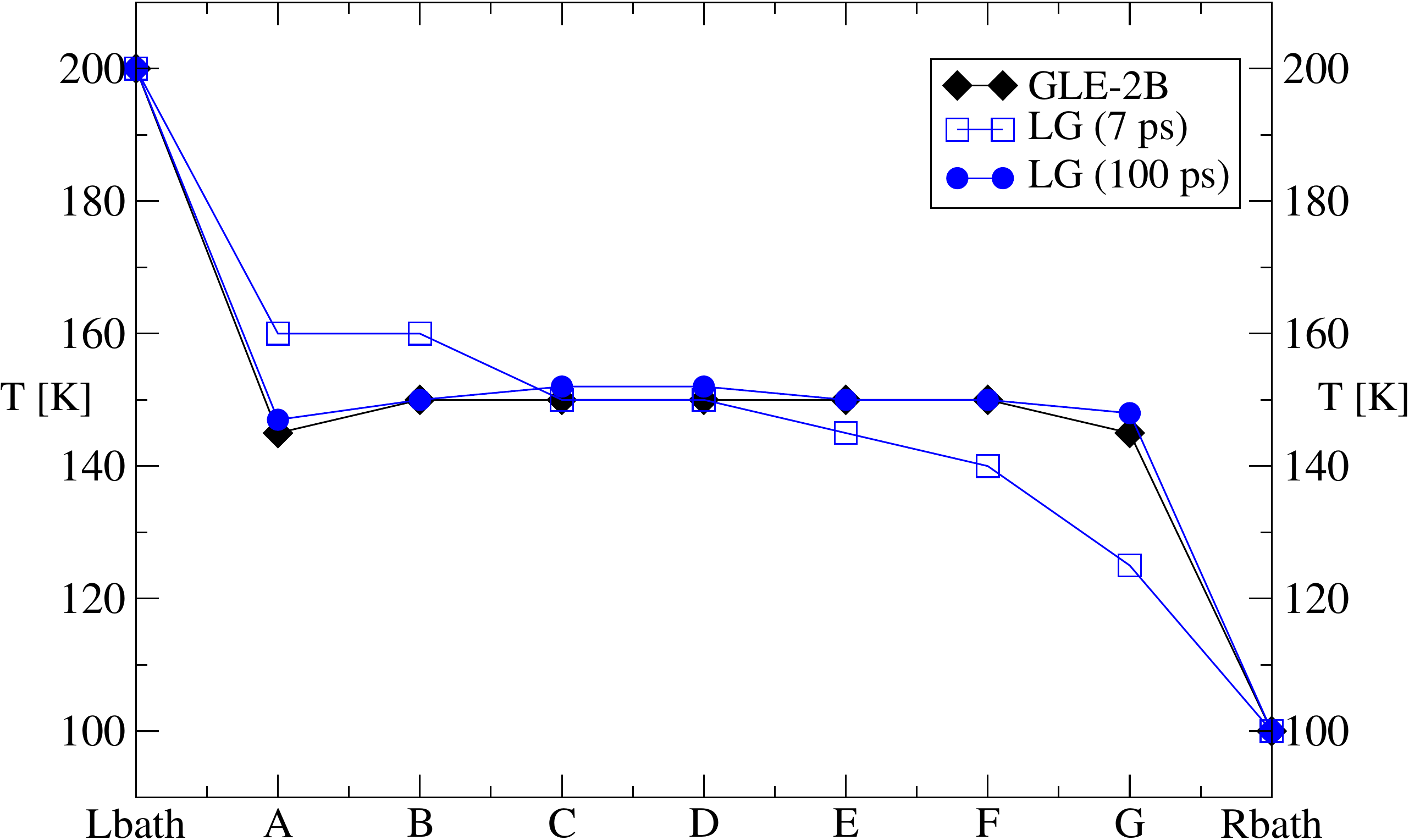}
\end{centering}
\caption{
Temperature profiles across the system consisting only of the 7 layers labelled A to G
in Figures \ref{fig:system_gle} and \ref{fig:GLE_and_otherthermostats}.
The different bath temperatures are $T_L=200$, $T_R=100$. 
While the GLE-2B calculations provide a uniform temperature profile inside the system, 
the LG thermostats show either the building up of a temperature 
gradient in the central system ($\tau_{\rm damp}=7$ ps) or a 
flat temperature profile ($\tau_{\rm damp}=100$ ps).
}
\label{fig:Tprofile_Alwire2}
\end{figure}

For this sytem, we obtain the same physical results for the temperature profile
as those presented in the main text [\onlinecite{Note:calcAlwire2}].
For low temperatures and small $\Delta T = T_L- T_R$, the central system is harmonic
and no temperature gradient is obtained across the system, see Figure \ref{fig:Tprofile_Alwire2}.

Once more, the LG thermostats provide two different temperature profiles depending upon
the value of the damping parameter $\tau_{\rm damp}$.
For small values of the damping parameter ($\tau_{\rm damp}=7$ ps), 
we obtain a small temperature gradient in the central system while for larger values of 
the damping parameter $\tau_{\rm damp}=100$ ps, the LG thermostats provide a flat 
temperature profile.

In terms of convergence versus the baths size, 
one should first note that, in the GLE-2B approach, we do not choose which group of atoms
are connected to a thermal baths, as this is usually performed when using more
conventional LG or NH thermostats.

The coupling of the central system to the thermal bath $\nu$
is obtained by the coupling of the DOF $i\alpha$ to the aDOF $s_{\nu,1}^{(k_\nu)}$
via the matrix elements $g_{i\alpha,b_\nu}$, see Eq.~(\ref{eq:gle_pia_Nbath})
and Eq.~(\ref{eq:extended_GLE}).
Such a coupling exists only when the matrix element $g_{i\alpha,b_\nu}$ is non-zero.
We recall that 
$g_{i\alpha,b_\nu}(\mathbf{r})=\partial f_{b_\nu}({\bf r})/\partial r_{i\alpha}$
is the derivative, with respect to the coordinate of the DOF $i\alpha$,
of the force $f_{b_\nu}$ felt by the atom $l_\nu$ of the bath $\nu$ with DOF $b_\nu \equiv l_\nu \gamma$.

The range of the matrix elements $g_{i\alpha,b_\nu}$ is determined by the cut-off 
of the inter-atomic potential used in the calculations. 
The range of the quantities $g_{i\alpha,b_\nu}$
is actually smaller than the cut-off of the inter-atomic potential, as the former 
is the second derivative (versus spatial coordinates) of the latter.

Hence there is no need to increase the size of the reduced bath region, as long as 
DOFs of the central system are properly coupled to the existing atoms in the reduced
bath region. Indeed, adding atoms in the reduced bath region (which corresponds
to $g_{i\alpha,b_\nu}=0$ elements) will not change the dissipative dynamics
of the atoms of the central region.

However, one should not forget that the infinite spatial extension of the baths has 
already been taken into account. 
In particular, the continuous vibrational spectra of the
infinite baths has been obtained through the calculation of their dynamical matrix
and subsequently in the mapping procedure described in Sections \ref{sec:GLEvDOF} 
and \ref{sec:calc_PI_mat} (for more detail, see also Ref.~[\onlinecite{Ness:2015}]).

The coupling to the thermal baths comes directly from the construction of the
geometry of the system itself and from the cut-off of the inter-atomic potential. 
It is not controlled by the user as usually done with LG or NH thermostats.
This is, once more, one of the main differences between the consistent (and more
elaborate) GLE-2B approach and the main-stream LG or NH thermostatting approach.

As far as the convergence versus the size of the central system is concerned, 
we have already mentioned in the main text that nonballistic transport properties
can be obtained for much longer and heterogeneous systems
\cite{Jackson:1968,Bolsterli:1970,Nakazawa:1970,Rubin:1971,Rich:1975,Spohn:1977,Davis:1978,
Kipnis:1982,Zurcher:1990,Saito:1996,Eckmann:1999,Hatano:1999,Tsironis:1999,Hu:2000,
Zhang:2002,Pereira:2004,Bernardin:2005,Mai:2006,Bricmont:2007,
Gaul:2007,Segal:2008,Lepri:2009,Hu:2010,Giberti:2011,Pereira:2011,Bernardin:2012,
Kannan:2012,Saaskilahti:2012,Shah:2013,Landi:2013}.
The presence of long wavelength accoustic modes and their indirect coupling (via
the baths) with other delocalised vibrational modes can also lead to the
establishment of a more diffusive transport property associated with the
presence of a temperature profile in the central system.
The study of the transport properties versus the size of the central system is 
important, but out of the scope of the present paper, and will be treated elsewhere.

\section{A one-dimensional toy model}
\label{app:1dmodel}

\begin{figure}
\begin{centering}
\includegraphics[width=70mm]{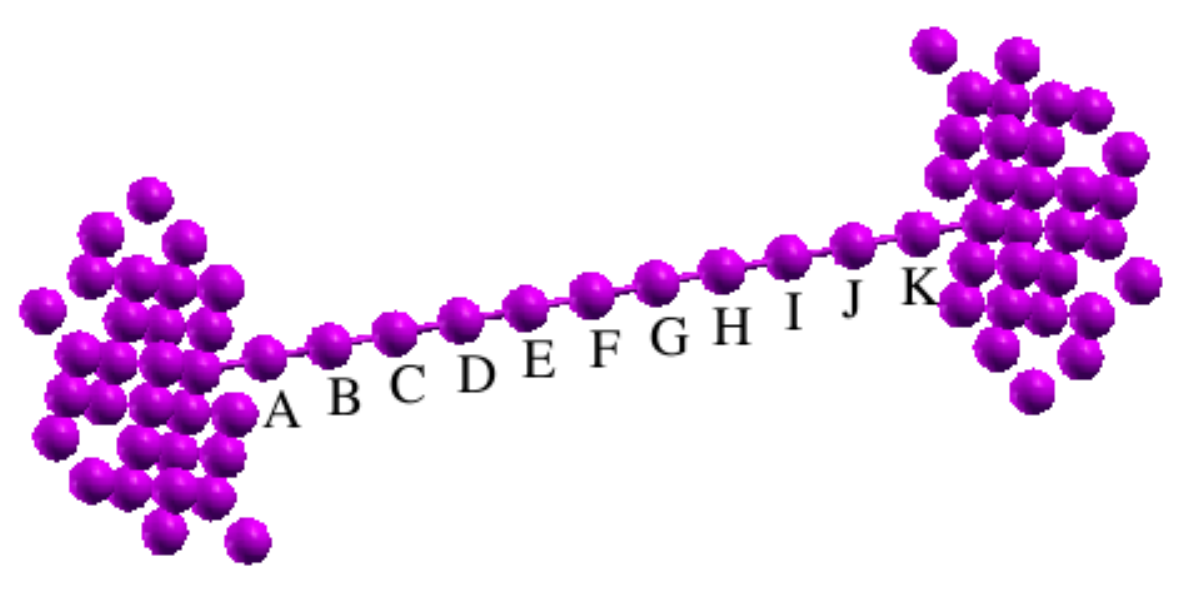}
\end{centering}
\caption{(Colour online)  
Model for a one-dimensional Al chain (central system containing 11 atoms labelled A to K) connected
to the left ($L$) and right ($R$) baths. For the GLE calculations, the $L$ and $R$ bath 
reduced region contains 30 atoms each. 
For the calculations of the dynamical matrix and the mapping of the polarisation matrix, 
larger baths (containing each 95 atoms) were considered.
The mapping of the polarisation matrix elements is done using 78 aDOF with
a set of 78 parameters $\{c_{b_\nu}^{(k_\nu)},\tau_{k_\nu},\omega_{k_\nu}\}$ for each bath
$\nu=L,R$.
}
\label{fig:1Dsystem}
\end{figure}

In this section, we consider a toy model for a one-dimensional system: a chain made
of 11 Al atoms connected to two baths as shown in Fig.~\ref{fig:1Dsystem}.
This is a simpler system than the three-dimensional wire considered in the main body
of the paper. We use the GLE-2B approach for the dynamics of the central chain
and study the local temperature profile of the perfect chain and of the chains
containing one or two defects. To model the defect, we simply change the mass
of the corresponding atom while conserving the same EAM inter-atomic potential for
all atoms in the GLE-2B calculations.

To make the system simpler, we constrain the dynamics of the atoms in the
central chain to a purely one-dimensional problem, i.e. the atoms can move only
along the main axis (the $x$-axis) of the chain. Therefore each atom in the central
region is associated with only one degree of freedom.

For a system at equilibrium, $T_L=T_R$, the GLE-2B calculations provide a velocity 
distribution for each atom of the central region which follows, as expected, 
the statistical Gaussian distribution of a single degree of freedom
given by 
\begin{equation}
\begin{split}
f_{\rm 1D}(v_x) = \sqrt{\frac{m}{2\pi kT}} \exp \left( - \frac{mv_x^2}{2kT} \right) \ .
\end{split}
\end{equation}

For the non-equilibrium conditions, $T_L \ne T_R$, the velocity distributions, 
calculated from the GLE-2B, for the atoms of the central chain also follow 
the lineshape of a Gaussian distribution. However the associated temperature
varies across the chain.

\begin{figure}
\begin{centering}
\includegraphics[width=75mm]{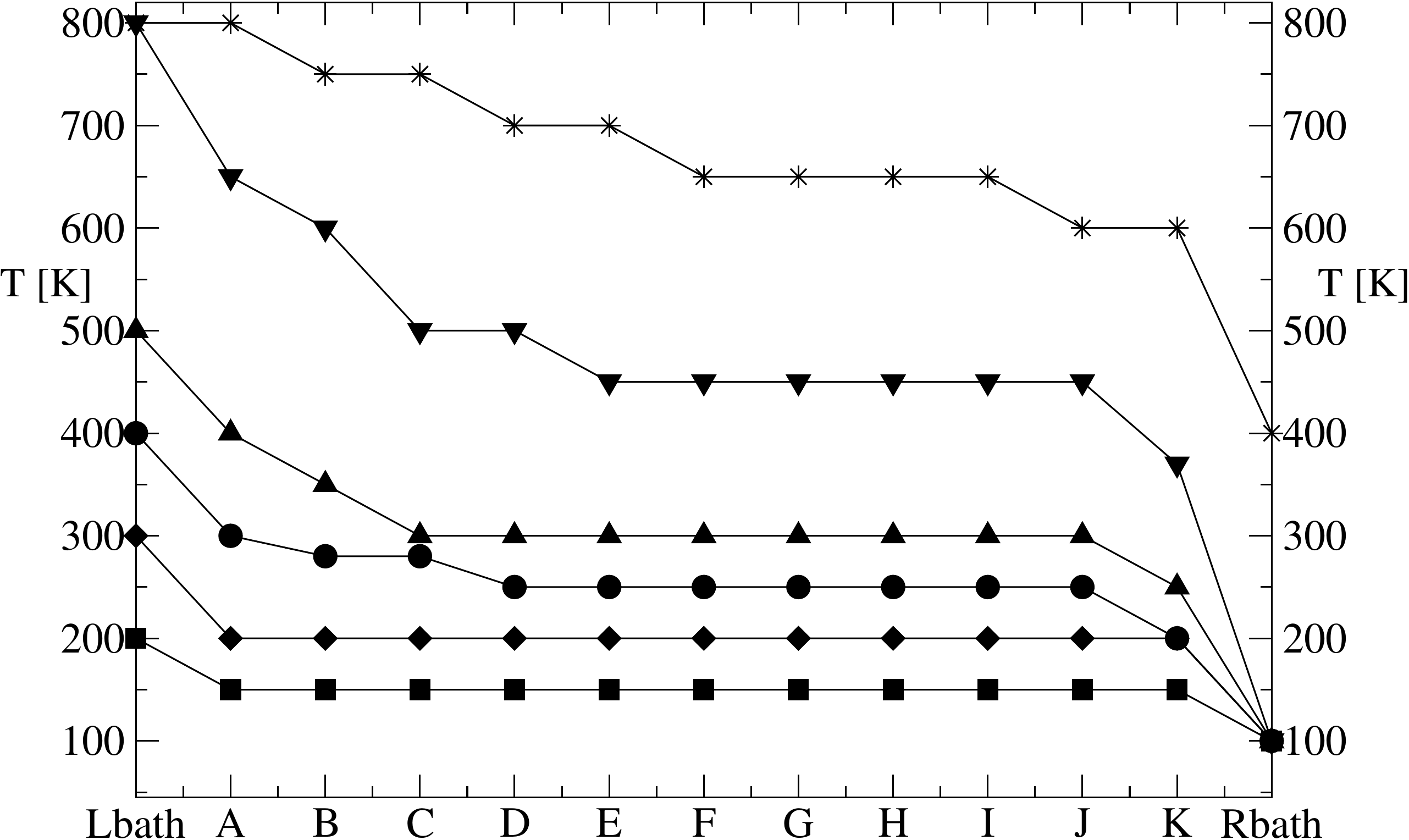}
\end{centering}
\caption{
Temperature profiles across the central 1D chain for different bath temperatures $T_{L,R}$. 
The labels Lbath (Rbath) represent the temperatures 
at the left (right) bath.
The other labels A to K indicate different atoms in the central chain. 
For ''low'' temperatures $T_{L,R}$, the system is harmonic and one gets a flat
temperature profile as expected for integrable systems.
For ''higher'' temperatures, the motion of some of the atoms sample the anharmonic part of
the interatomic potential, and therefore the system is non-integrable and a 
temperature gradient is built up.
}
\label{fig:Tprofile_1Dchain}
\end{figure}

Figure~\ref{fig:Tprofile_1Dchain} shows the temperature profile across the
chain when the masses of all atoms in the central chain are equal.
For ''low'' temperatures $T_{L,R}$, the homogeneous system behaves as a harmonic 
system and one obtains a flat temperature profile as expected for integrable systems.
For ''higher'' temperatures, the motion of some of the atoms start to sample the 
anharmonic part of the inter-atomic potential. Therefore some parts of the system are
not harmonic any more and a temperature gradient starts to build up.

Note that because the dynamics of the system is strongly constrained, no
structural instability is possible and higher local effective temperatures (in comparison
with the 3D short wire considered in the main part of the paper) can be investigated
in order to achieve the anharmonic regime.
  
\begin{figure}
\begin{centering}
\includegraphics[width=75mm]{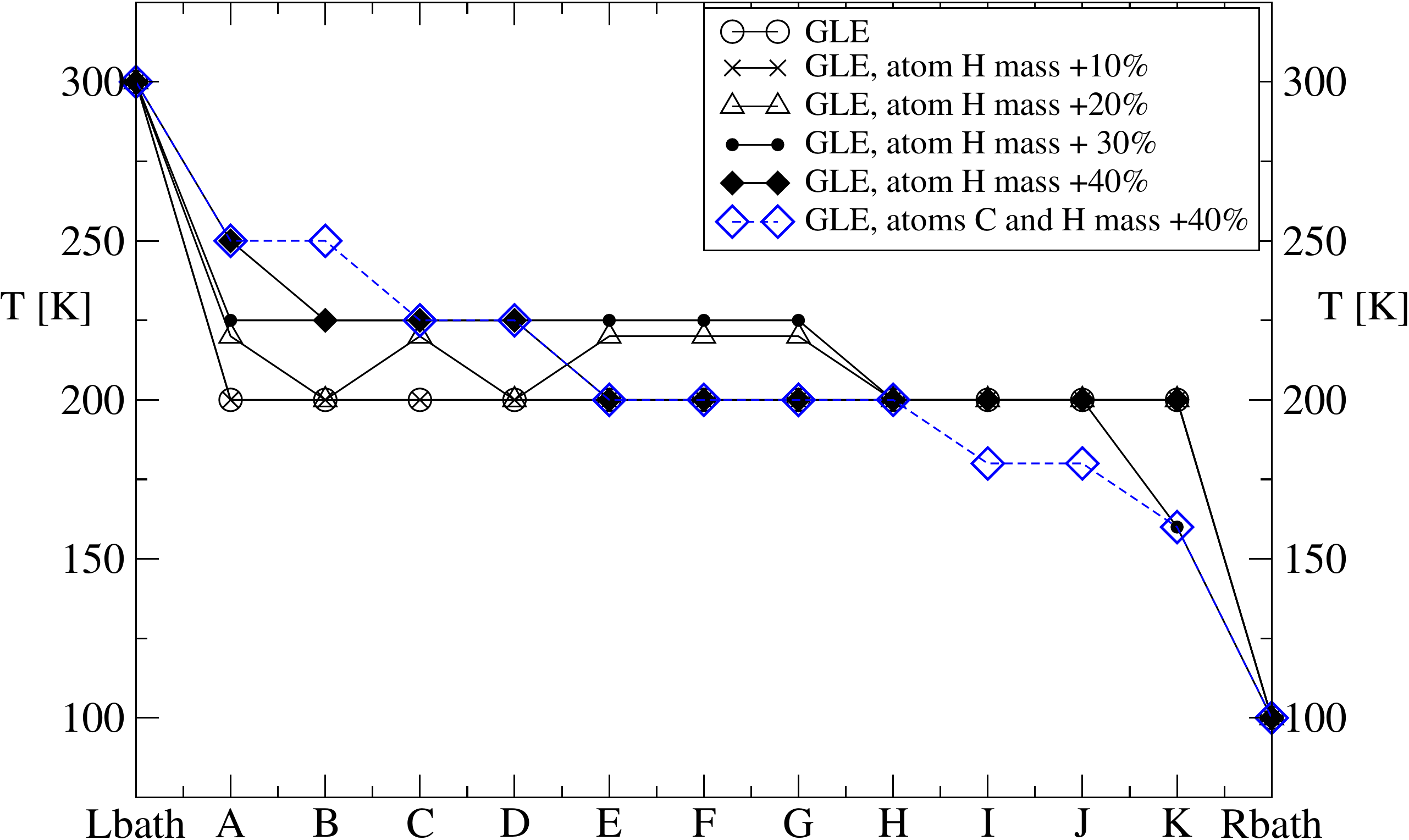}
\end{centering}
\caption{(Colour online)  
Temperature profiles across the central 1D chain for different defects present in the chain. 
The labels Lbath (Rbath) represent the temperatures 
at the left (right) bath.
The other labels A to K indicate different atoms in the central chain. 
The mass of atom H in Fig.~\ref{fig:1Dsystem} is increased by 10, 20, 30 or 40\%. The case
of two defects (atoms C and H with their mass increased by 40\%) is also shown.
For the temperatures considered, the chain is in the harmonic regime. However the presence
of defects, and their associated more localised vibration modes, leads to the build up of
a temperature gradient across the chain.
}
\label{fig:Tprofile_1Dchain_chmass}
\end{figure}

Figure~\ref{fig:Tprofile_1Dchain_chmass} shows the temperature profile across the
chain when one introduces a localised defect in the chain. Calculations are performed
when the mass of the atom labelled H in the chain is increased by 10 to 40\% and
when the mass of both atoms C and H is increased by 40\%.  
One can see that the introduction of a defect in the chain (in the harmonic regime)
leads to the build-up of a temperature gradient. 
Such an effect is clearly obtained for a mass increase larger than 10~\% and in the 
cases of more than one defect present in the chain.
 
We have checked that, in the presence of defects, the vibration modes of the chain are
slightly more localised (around the defects) than for the perfect chain. Furthermore
the amplitudes of the vibration modes at the ends of the chain are also different in
the presence of defects in the chain. Such effects are thought to hinder the heat transport
from one bath to the other and hence lead to the absence of a flat temperature
profile across the chain.

Thus, the calculations shown here for a toy model of a one-dimensional Al chain
present the same qualitative physics as that obtained for the three-dimensional short
wire described in the main text.


\begin{thebibliography}{255}

\expandafter\ifx\csname natexlab\endcsname\relax\def\natexlab#1{#1}\fi
\expandafter\ifx\csname bibnamefont\endcsname\relax
  \def\bibnamefont#1{#1}\fi
\expandafter\ifx\csname bibfnamefont\endcsname\relax
  \def\bibfnamefont#1{#1}\fi
\expandafter\ifx\csname citenamefont\endcsname\relax
  \def\citenamefont#1{#1}\fi
\expandafter\ifx\csname url\endcsname\relax
  \def\url#1{\texttt{#1}}\fi
\expandafter\ifx\csname urlprefix\endcsname\relax\def\urlprefix{URL }\fi
\providecommand{\bibinfo}[2]{#2}
\providecommand{\eprint}[2][]{\url{#2}}

\bibitem{Berber2000} 
  \bibinfo {author} {\bibfnamefont{S.}~\bibnamefont{Berber}}, \bibinfo {author}
  {\bibfnamefont{Y.-K.}\ \bibnamefont{Kwon}},\ and\ \bibinfo {author}
  {\bibfnamefont{D.}~\bibnamefont{Tom\'{a}nek}},\ 
  \bibinfo {journal} {Phys. Rev. Lett.}\ 
  \textbf{\bibinfo {volume} {84}},\ \bibinfo {pages} {4613} (\bibinfo {year}
  {2000})%

\bibitem{Kim2001}
  \bibinfo {author} {\bibfnamefont{P.}~\bibnamefont{Kim}}, \bibinfo {author}
  {\bibfnamefont{L.}~\bibnamefont{Shi}}, \bibinfo {author}
  {\bibfnamefont{A.}~\bibnamefont{Majumdar}},\ and\ \bibinfo {author}
  {\bibfnamefont{P.~L.}\ \bibnamefont{McEuen}},\  
  \bibinfo {journal} {Phys. Rev. Lett.}\  
  \textbf{\bibinfo {volume} {87}},\ \bibinfo {pages} {215502} (\bibinfo {year}
  {2001})%

\bibitem{Shi2002} 
  \bibinfo {author} {\bibfnamefont{L.}~\bibnamefont{Shi}}\ and\ \bibinfo
  {author} {\bibfnamefont{A.}~\bibnamefont{Majumdar}},\  
  \bibinfo {journal} {J. Heat Trans. - T. ASME}\  
  \textbf{\bibinfo {volume} {124}},\ \bibinfo {pages} {329} (\bibinfo {year}
  {2002})%

\bibitem{Padgett2004} 
  \bibinfo {author} {\bibfnamefont{C.~W.}\ \bibnamefont{Padgett}}\ and\
  \bibinfo {author} {\bibfnamefont{D.~W.}\ \bibnamefont{Brenner}},\  
  \bibinfo {journal} {Nano Letters}\  
  \textbf{\bibinfo {volume} {4}},\ \bibinfo {pages} {1051} (\bibinfo {year}
  {2004})%

\bibitem{Hu2008} 
  \bibinfo {author} {\bibfnamefont{M.}~\bibnamefont{Hu}}, \bibinfo {author}
  {\bibfnamefont{P.}~\bibnamefont{Keblinski}}, \bibinfo {author}
  {\bibfnamefont{J.-S.}\ \bibnamefont{Wang}},\ and\ \bibinfo {author}
  {\bibfnamefont{N.}~\bibnamefont{Raravikar}},\  
  \bibinfo {journal} {Journal of Applied Physics}\  
  \textbf{\bibinfo {volume} {104}},\ \bibinfo {pages} {083503} (\bibinfo {year}
  {2008})%

\bibitem{Padgett2006} 
  \bibinfo {author} {\bibfnamefont{C.~W.}\ \bibnamefont{Padgett}}, \bibinfo
  {author} {\bibfnamefont{O.}~\bibnamefont{Shenderova}},\ and\ \bibinfo
  {author} {\bibfnamefont{D.~W.}\ \bibnamefont{Brenner}},\  
  \bibinfo {journal} {Nano Lett.}\  
  \textbf{\bibinfo {volume} {6}},\ \bibinfo {pages} {1827} (\bibinfo {year}
  {2006}).%

\bibitem{Yang2008} 
  \bibinfo {author} {\bibfnamefont{N.}~\bibnamefont{Yang}}, \bibinfo {author}
  {\bibfnamefont{G.}~\bibnamefont{Zhang}},\ and\ \bibinfo {author}
  {\bibfnamefont{B.}~\bibnamefont{Li}},\  
  \bibinfo {journal} {Nano Lett.}\  
  \textbf{\bibinfo {volume} {8}},\ \bibinfo {pages} {276} (\bibinfo {year}
  {2008}).%

\bibitem{Estreicher2009} 
  \bibinfo {author} {\bibfnamefont{S.~K.}\ \bibnamefont{Estreicher}}\ and\
  \bibinfo {author} {\bibfnamefont{T.~M.}\ \bibnamefont{Gibbons}},\  
  \bibinfo {journal} {Physica B}\  
  \textbf{\bibinfo {volume} {404}},\ \bibinfo {pages} {4509} (\bibinfo {year}
  {2009}).%

\bibitem{Majumdar:1999}
A. Majumdar, 
Annu. Rev. Mater. Sci. {\bf 29}, 505 (1999).

\bibitem{Segal2002} 
  \bibinfo {author} {\bibfnamefont{D.}~\bibnamefont{Segal}}\ and\ \bibinfo{author} 
 {\bibfnamefont{A.}~\bibnamefont{Nitzan}},\ 
  \bibinfo {journal} {J. Chem. Phys.}\ 
  \textbf{\bibinfo {volume} {117}},\ \bibinfo {pages} {3915} (\bibinfo {year}
  {2002}).%

\bibitem{Segal:2003}
D. Segal, A. Nitzan and P. H\"anggi, J. Chem. Phys. {\bf 119}, 6840 (2003).


\bibitem{Mingo:2003}
N. Mingo and Liu Yang, Phys. Rev. B {\bf 68}, 245406 (2003).

\bibitem{Yao:2005}
Z. Yao, J.-S. Wang, B. Li and G.-R. Liu, Phys. Rev. B {\bf 71}, 085417 (2005).

\bibitem{Wang:2007}
J.-S. Wang, Phys. Rev. Lett. {\bf 99}, 160601 (2007).

\bibitem{Dubi2011} 
  \bibinfo {author} {\bibfnamefont{Y.}~\bibnamefont{Dubi}}\ and\ \bibinfo
  {author} {\bibfnamefont{M.}~\bibnamefont{Di~Ventra}},\
  \bibinfo {journal} {Rev. Mod. Phys.}\
  \textbf{\bibinfo {volume} {83}},\ \bibinfo {pages} {131} (\bibinfo {year}
  {2011}).%

\bibitem{Widawsky:2012}
J. R. Widawsky, P. Darancet, J. B. Neaton, and L. Venkataraman,
Nano Lett. 12, 354 (2012).

\bibitem{Cahill2002} 
  \bibinfo {author} {\bibfnamefont{D.~G.}\ \bibnamefont{Cahill}}, \bibinfo
  {author} {\bibfnamefont{K.}~\bibnamefont{Goodson}},\ and\ \bibinfo {author}
  {\bibfnamefont{A.}~\bibnamefont{Majumdar}},\  
  \bibinfo {journal} {J. Heat Trans. - T. ASME}\  
  \textbf{\bibinfo {volume} {124}},\ \bibinfo {pages} {223} (\bibinfo {year}
  {2002})%

\bibitem{Pop2010} 
  \bibinfo {author} {\bibfnamefont{E.}~\bibnamefont{Pop}},\  
  \bibinfo {journal} {Nano. Res.}\  
  \textbf{\bibinfo {volume} {3}},\ \bibinfo {pages} {147} (\bibinfo {year}
  {2010})%

\bibitem{Zebarjadi2012} 
  \bibinfo {author} {\bibfnamefont{M.}~\bibnamefont{Zebarjadi}}, \bibinfo
  {author} {\bibfnamefont{K.}~\bibnamefont{Esfarjani}}, \bibinfo {author}
  {\bibfnamefont{M.~S.}\ \bibnamefont{Dresselhaus}}, \bibinfo {author}
  {\bibfnamefont{Z.~F.}\ \bibnamefont{Ren}},\ and\ \bibinfo {author}
  {\bibfnamefont{G.}~\bibnamefont{Chen}},\  
  \bibinfo {journal} {Energy Environ. Sci.}\  
  \textbf{\bibinfo {volume} {5}},\ \bibinfo {pages} {5147} (\bibinfo {year}
  {2012})%

\bibitem{Gotsmann:2013}
B. Gotsmann, M. A. Lantz,
Nature Materials {\bf 12}, 59 (2013).

\bibitem{Menges:2013}
F. Menges, H. Riel, A. Stemmer, C. Dimitrakopoulos and B. Gotsmann,
Phys. Rev. Lett. {\bf 111}, 205901 (2013).

\bibitem{Meier:2014}
T. Meier, F. Menges, P. Nirmalraj, H. H\"olscher, H. Riel and B. Gotsmann
Phys. Rev. Lett. {\bf 113}, 060801 (2014).

\bibitem{Li:2015}
J. Li, D.-Q. Zheng and W.-R. Zhong,
EuroPhys. Lett. {\bf 112}, 24006 (2015).


\bibitem{Mori:1965}
H. Mori, Prog. Theor. Phys. {\bf 33}, 423 (1965).

\bibitem{Adelman:1976}
S. A. Adelman and J. Doll, J. Chem. Phys. {\bf 64}, 2375 (1976).

\bibitem{Adelman:1980}
S. A. Adelman, J. Chem. Phys. {\bf 73}, 3145 (1980).

\bibitem{Ermak:1980}
D. L. Ermak and H. Buckholz, J. Comp. Phys. {\bf 35}, 169 (1980).

\bibitem{Carmeli:1983}
B. Carmeli and A. Nitzan, Chem. Phys. Lett. {\bf 102}, 517 (1983).

\bibitem{Cortes:1985}
E. Cort\'es, B. J. West and K. Lindenberg, J. Chem. Phys. {\bf 82}, 2708 (1985).

\bibitem{Lindenberg:1990}
K. Lindenberg and B. J. West,
{\it The Nonequilibrium Statistical Mechanics of Open and Closed Systems}
(Wiley-VCH, New York, 1990).

\bibitem{Tsekov:1994a}
R. Tsekov and E. Ruckenstein, J. Chem. Phys. {\bf 100}, 1450 (1994).

\bibitem{Tsekov:1994b}
R. Tsekov and E. Ruckenstein, J. Chem. Phys. {\bf 101}, 7844 (1994).

\bibitem{Risken:1996}
H. Risken, {\it The Fokker-Planck Equation: Methods of Solutions
and Applications, 2nd ed.} (Springer, Berlin, 1996).

\bibitem{Hernandez:1999}
R. Hernandez, J. Chem. Phys. {\bf 111}, 7701 (1999).

\bibitem{Zwanzig:2001} 
  \bibinfo {author} {\bibfnamefont{R.}~\bibnamefont{Zwanzig}},\  
  \emph{\bibinfo {title} {Nonequilibrium Statistical Mechanics}}\ (\bibinfo
  {publisher} {Oxford University Press},\ \bibinfo {year} {2001})%

\bibitem{Kupferman:2004}
R. Kupferman, J. Stat. Phys. {\bf 114}, 291 (2004). 

\bibitem{Bao:2004}
J.-D. Bao, J. Stat. Phys. {\bf 114}, 503 (2004).

\bibitem{Luczka:2005}
J. {\L}uczka, Chaos {\bf 15}, 026107 (2005).

\bibitem{Izvekov:2006}
S. Izvekov and G. A. Voth, J. Chem. Phys. {\bf 125}, 151101 (2006).

\bibitem{Snook:2007}
I. Snook, {\it The Langevin and Generalised Langevin Approach to the Dynamics 
of Atomic, Polymeric and Colloidal Systems} (Elsevier, Amsterdam, 2007).

\bibitem{vanVliet:2008}
C. M. van Vliet, 
{\it Equilibrium and Non-Equilibrium Statistical Mechanics}
(World Scientific, Singapore, 2008).

\bibitem{Kantorovich:2008}
L. Kantorovich, Phys. Rev. B {\bf 78} 094304 (2008).

\bibitem{Ceriotti:2009}
M. Ceriotti, G. Bussi and M. Parrinello, Phys. Rev. Lett. {\bf 102}, 020601 (2009).

\bibitem[{\citenamefont{Ceriotti et~al.}(2010)\citenamefont{Ceriotti, Bussi,
  and Parrinello}}]{Ceriotti:2010}
\bibinfo{author}{\bibfnamefont{M.}~\bibnamefont{Ceriotti}},
  \bibinfo{author}{\bibfnamefont{G.}~\bibnamefont{Bussi}}, \bibnamefont{and}
  \bibinfo{author}{\bibfnamefont{M.}~\bibnamefont{Parrinello}},
  \bibinfo{journal}{J. Chem. Theory Comput.} \textbf{\bibinfo{volume}{6}},
  \bibinfo{pages}{1170} (\bibinfo{year}{2010}).

\bibitem{Siegle:2010}
P. Siegle, I. Goychuk, P. Talkner and Peter H\"anggi, Phys. Rev. E {\bf 81}, 011136 (2010).

\bibitem{Kawai:2011}
S. Kawai and T. Komatsuzaki, J. Chem. Phys. {\bf 134}, 114523 (2011).

\bibitem[{\citenamefont{Morrone et~al.}(2011)\citenamefont{Morrone, Markland,
  Ceriotti, and Berne}}]{Morrone:2011}
\bibinfo{author}{\bibfnamefont{J.~A.} \bibnamefont{Morrone}},
  \bibinfo{author}{\bibfnamefont{T.~E.} \bibnamefont{Markland}},
  \bibinfo{author}{\bibfnamefont{M.}~\bibnamefont{Ceriotti}}, \bibnamefont{and}
  \bibinfo{author}{\bibfnamefont{B.~J.} \bibnamefont{Berne}},
  \bibinfo{journal}{J. Chem. Phys.} \textbf{\bibinfo{volume}{134}},
  \bibinfo{pages}{014103} (\bibinfo{year}{2011}).

\bibitem[{\citenamefont{Ceriotti et~al.}(2011)\citenamefont{Ceriotti,
  Manolopoulos, and Parrinello}}]{Ceriotti:2011}
\bibinfo{author}{\bibfnamefont{M.}~\bibnamefont{Ceriotti}},
  \bibinfo{author}{\bibfnamefont{D.~E.} \bibnamefont{Manolopoulos}},
  \bibnamefont{and}
  \bibinfo{author}{\bibfnamefont{M.}~\bibnamefont{Parrinello}},
  \bibinfo{journal}{J. Chem. Phys.} \textbf{\bibinfo{volume}{134}},
  \bibinfo{pages}{084104} (\bibinfo{year}{2011}).

\bibitem{Pagel:2013}
D. Pagel, A. Alvermann and H. Fehske, Phys. Rev. E {\bf 87}, 012127 (2013).

\bibitem{Leimkuhler:2013}
B. Leimkuhler and C. Matthews, J. Chem. Phys. {\bf 138}, 174102 (2013).

\bibitem{Baczewski:2013}
A. D. Baczewski and S. D. Bond, J. Chem. Phys. {\bf 139}, 044107 (2013).

\bibitem[{\citenamefont{Stella et~al.}(2014)\citenamefont{Stella, Lorenz, and
  Kantorovich}}]{Stella:2014}
\bibinfo{author}{\bibfnamefont{L.}~\bibnamefont{Stella}},
  \bibinfo{author}{\bibfnamefont{C.~D.} \bibnamefont{Lorenz}},
  \bibnamefont{and}
  \bibinfo{author}{\bibfnamefont{L.}~\bibnamefont{Kantorovich}},
  \bibinfo{journal}{Physical Review B} \textbf{\bibinfo{volume}{89}},
  \bibinfo{pages}{134303} (\bibinfo{year}{2014}).

\bibitem{Ness:2015}
H. Ness, L. Stella, C. D. Lorenz and L. Kantorovich, Phys. Rev. B {\bf 91}, 014301 (2015).

\bibitem{Andersen:1980}
H. C. Andersen, J. Chem. Phys. {\bf 72}, 2384 (1980).

\bibitem{Nose:1984a}
S. Nos\'e, Mol. Phys. {\bf 52}, 255 (1984).

\bibitem{Nose:1984b}
S. Nos\'e, J. Chem. Phys. {\bf 81}, 511 (1984).

\bibitem{Hoover:1985}
W. G. Hoover, Phys. Rev. A {\bf 31}, 1695 (1985).

\bibitem{Toton:2010}
D. Toton, C. D. Lorenz, N. Rompotis, N. Martsinovich, and L. Kantorovich, 
J. Phys.: Condens. Matter {\bf 22}, 074205 (2010).

\bibitem{Kantorovich:2008b}
L. Kantorovich and N Rompotis, Phys. Rev. B {\bf 78} 094305 (2008).

\bibitem{Plimpton:1995}
S. J. Plimpton, J. Comput. Phys. {\bf 117}, 1 (1995)

\bibitem{Gillespie:1996b}
D. T. Gillespie,
Am. J. Phys. {\bf 64}, 1246 (1996).

\bibitem{Gillespie:1996}
D. T. Gillespie,
Am. J. Phys. {\bf 64}, 225 (1996).

\bibitem{Note:trotter_decomp}
There exists different possible choice for the Trotter factorization
of the Liouvillian.
From our understanding of Ref.~[\onlinecite{Leimkuhler:2013}], 
the Trotter decomposition we use to
obtain our GLE-MD algorithm has been shown to provide a more 
appropriate description of the velocity correlation functions.
An accurate statistical description of the velocities is
crucial in the present paper, as we extract the local temperature
from the velocities statistical distributions.

\bibitem{Note:1}
It should noted that there cannot be any form of coupling between the two baths. 
Otherwise they would not be independent and therefore it would not be possible to consider
them as being kept at their own thermodynamic equilibrium
characterised by the temperature $T_\nu$.

\bibitem{Ferrario:1979}
M. Ferrario and P. Grigolini, J. Math. Phys. {\bf 20}, 2567 (1979).

\bibitem{Marchesoni:1983}
F. Marchesoni and P. Grigolini, J. Chem. Phys. {\bf 78}, 6287 (1983).


\bibitem{Wan:1997}
C. C. Wan, J.-L. Mozos, J. Wang and H. Guo,
Phys. Rev. B {\bf 55}, R13393(R) (1997).

\bibitem{Taraschi:1998}
G. Taraschi, J.-L. Mozos, C. C. Wan, H. Guo and J. Wang
Phys. Rev. B {\bf 58}, 13138 (1998).

\bibitem{Taylor:2001}
J. Taylor, H. Guo and J. Wang,
Phys. Rev. B {\bf 63}, 245407 (2001).

\bibitem{Daw:1984}
M. S. Daw and M. I. Baskes, 
Phys. Rev. B {\bf 29}, 6443 (1984).

\bibitem{NISTweb}
We have use the data file Al99.eam.alloy from the NIST Interatomic Potential Repository Project
(http://www.ctcms.nist.gov/potentials/) as provided by Ref.~[\onlinecite{Mishin:1999}].

\bibitem{Mishin:1999}
Y. Mishin, D. Farkas, M.J. Mehl, and D.A. Papaconstantopoulos, 
Phys. Rev. B 59, 3393 (1999). 

\bibitem{Note:2} The velocity distributions are calculated from the averages of several
realisations of each velocity distribution obtained at a given timestep, after the
system is thermalised. For $T$ = 200 K, the thermalisation is
reached after 78 ps of GLE dynamics, i.e. after 40000 time-steps with
$\Delta t$ = 2 fs. The distributions are averaged over all the time-steps 
in the time range $t=[78,144]$ ps. 

\bibitem{Note:melting}
Calculation using the LG thermostats and the GLE-2B approach
were performed at equilibrium when the system is uniformly 
thermostatted at $T_L = T_R$. 
From the results of these calculations, we observed that
the Al nanowire melts down at T = 500 K for the EAM inter-atomic
potential we used. During the MD run, we observe the formation
of the smaller constriction in the middle of the nanowire and
eventually all the atoms of the layers A to F end up on the surfaces
of the L and R bath regions.
At temperatures T = 400 - 450, we also obtain the formation of 
a smaller constriction in the middle of the nanowire (with atoms 
moving towards the left and right sides of the constriction).
The constriction consists of 3 layers of only 2 atoms in each layer.
This is what we call "structural instabilities",
meaning that the central system does not have the same atomic 
configuration (layers made of 4 or 5 atoms).
Such strong modifications of the configuration of the system make it
difficult to perform a consistent analysis of the (layer-by-layer)
temperature profiles.

One should notice that the melting of Al surfaces, depending
on the Miller indices $(hkl)$, may occur even at temperatures which are
150 to 200 K below the bulk melting temperature [see for example
P. Stoltze, J. K. N{\o}rskov, and U. Landman, Phys. Rev. Lett. {\bf 61}, 440 (1988) ], 
i.e. roughly around 730 - 780 K. 
The coordination of an Al atom at the surface is roughly half of that of 
an atom in the bulk. Therefore we believe that it is not surprising
that the melting of the nanowire happens at even lower temperature as
the coordination of the atoms inside the nanowire is even smaller that
the coordination of surface atoms.

\bibitem{Small:2003}
J. P. Small, L. Shi and P. Kim, 
Solid State Comm. {\bf 127}, 181 (2003).

\bibitem{Note:frozenmiddle}
The GLE-2B runs are performed for 300 ps 
(150000 timesteps with $\Delta t=2$ fs).
The velocity histograms are obtained
from statistical averaging over the time range 200 to 280 ps.



\bibitem{Rieder:1967}
Z. Rieder, J. L. Lebowitz and E. Lieb, 
J. Math. Phys. {\bf 8}, 1073 (1967).

\bibitem{Rich:1975}
M. Rich and W. M. Visscher, 
Phys. Rev. B {\bf 11}, 2164 (1975).

\bibitem{Spohn:1977}
H. Spohn and J. L. Lebowitz,
Commun. Math. Phys. {\bf 54}, 97 (1977).

\bibitem{Casati:1979}
G. Casati,
Nuovo Cimento {\bf 52}, 257 (1979).

\bibitem{Hu:2000}
B. Hu and B. Li and H. Zhao,
Phys. Rev. E {\bf 61}, 3828 (2000).

\bibitem{Segal:2008}
D. Segal,
J. Chem. Phys. {\bf 128}, 224710 (2008).

\bibitem{Kannan:2012}
V. Kannan, A. Dhar and J. L. Lebowitz,
Phys. Rev. E {\bf 85}, 041118 (2012).

\bibitem{Saaskilahti:2012}
K. S\"a\"askilahti, J. Oksanen, R. P. Linna and J. Tulkki,
Phys. Rev. E {\bf 86}, 031107 (2012).

\bibitem{Landi:2013}
G. T. Landi and M. J. de Oliveira,
Phys. Rev. E {\bf 87}, 052126 (2013).

\bibitem{Zurcher:1990}
U. Z\"urcher and P. Talkner,
Phys. Rev. A {\bf 42}, 3278 (1990).

\bibitem{Saito:1996}
K. Saito, S. Takesue and S. Miyashita,
Phys. Rev. E {\bf 54}, 2404 (1996).

\bibitem{Saito:2000}
K. Saito, S. Takesue and S. Miyashita,		
Phys. Rev. E {\bf 61}, 2397 (2000).

\bibitem{Dahr:2003}
A. Dahr and B. Sriram Shastry,
Phys. Rev. B {\bf 67}, 195405 (2003).

\bibitem{Gaul:2007}
C. Gaul and H. B\"uttner,
Phys. Rev. E {\bf 76}, 011111 (2007).

\bibitem{Asadian:2013}
A. Asadian, D. Manzano, M. Tiersch, and H. J. Briegel,
Phys. Rev. E {\bf 87}, 012109 (2013).

\bibitem{Jackson:1968}
E. A. Jackson. J. R. Pasta and J. F. Waters,
J. Comput. Phys. {\bf 2}, 207 (1968).

\bibitem{Rubin:1971}
R. J. Rubin and W. L. Greer,
J. Math. Phys. {\bf 12}, 1686 (1971).

\bibitem{Bolsterli:1970}
M. Bolsterli, M. Rich, and W. M. Visscher,
Phys. Rev. A {\bf 1}, 1086 (1970)

\bibitem{Nakazawa:1970}
H. Nakazawa, Prog. Theor. Phys. Suppl. {\bf 45}, 231 (1970).

\bibitem{Stoneham:1975}
A. M. Stoneham, 
{\it Theory of Defects in Solids: Electronic Structure of Defects 
in Insulators and Semiconductors}
(Oxford University Press, Oxford, 1975). Chapter 11.

\bibitem{Eckmann:1999}
J.-P. Eckmann, C.-A. Pillet and L. Rey-Bellet,
Commun. Math. Phys. {\bf 201}, 657 (1999).

\bibitem{Hatano:1999}
T. Hatano,	
Phys. Rev. E {\bf 59}, R1 (1999).

\bibitem{Zhang:2002}
Y. Zhang and H. Zhao,
Phys. Rev. E {\bf 66}, 026106 (2002).

\bibitem{Pereira:2004}
E. Pereira and R. Falcao,
Phys. Rev. E {\bf 70}, 046105 (2004).

\bibitem{Mai:2006}
T. Mai and O. Narayan,
Phys. Rev. E {\bf 73}, 061202 (2006).

\bibitem{Bricmont:2007}
J. Bricmont and A. Kupiainen,
Commun. Math. Phys. {\bf 274}, 555 (2007).

\bibitem{Hu:2010}
T. Hu and Y. Tang,
J.Phys.Soc.Jpn. {\bf 79}, 064601 (2010).

\bibitem{Giberti:2011}
C. Giberti and L. Rondoni,
Phys. Rev. E {\bf 83}, 041115 (2011).

\bibitem{Pereira:2011}
E. Pereira,		
Physica A {\bf 390}, 4131 (2011).

\bibitem{Shah:2013}
T.N. Shah and P.N. Gajjar,
Commun. Theor. Phys. {\bf 59}, 361 (2013).

\bibitem{Tsironis:1999}
G.P. Tsironis, A. R. Bishop, A. V. Savin and A. V. Zolotaryuk,
Phys. Rev. E {\bf 60}, 6610 (1999).

\bibitem{Kipnis:1982}
C. Kipnis, C. Marchioro and E. Presutti, 
J. Stat. Phys. {\bf 27}, 65 (1982).

\bibitem{Bernardin:2005}
C. Bernardin and S. Olla,
J. Stat. Phys. {\bf 121}, 271 (2005).

\bibitem{Lepri:2009}
S. Lepri, C. Mej\'{\i}a-Monasterio and A. Politi,
J. Math. A: Math. Theor. {\bf 42}, 025010 (2009).

\bibitem{Bernardin:2012}
C. Bernardin, V. Kannan, J.L. Lebowitz and J. Lukkarinen,
J. Stat. Phys. {\bf 146}, 800 (2012).

\bibitem{Davis:1978}
E.B. Davis,
J. Stat. Phys. {\bf 18}, 161 (1978).

\bibitem{Wang:2004}
J.-S. Wang and B. Li,
Phys. Rev. E {\bf 70}, 021204 (2004).

\bibitem{Ceriotti:2009b}
M. Ceriotti, G. Bussi and M. Parrinello,
Phys. Rev. Lett. {\bf 103}, 030603 (2009).

\bibitem{Note:enerpot}
The atoms of the bath reduced regions are present in the GLE-2B method (implemented in
the LAMMPS code), as one needs to calculate the matrix elements $g_{i\alpha,b_\nu}(\mathbf{r})$
for the instantaneous atomic positions of the central system.
However the atoms in the bath reduced regions are kept fixed during the GLE-2B runs,
hence they do not contribute to the total kinetic energy of the system shown in
Fig.~\ref{fig:system}. Futhermore the contribution, to the total potential
energy, from these atoms is just a constant throughout the GLE-2B runs. 

\bibitem{Note:calcAlwire2}
The GLE-2B runs, as well as the LG runs with $\tau_{\rm damp}=7$ ps, are performed for 120 ps 
(120000 timesteps with $\Delta t=1$ fs).
The steady state is reached around 80 ps.
The velocity histograms (from which the local temperature is extracted) are obtained
from statistical averaging over the time range 100 to 120 ps.
For the LG runs with $\tau_{\rm damp}=100$ ps, the steady state is reached after a longer
time, $\sim$ 300 ps.  The velocity histograms are obtained
from statistical averaging over the time range 450 to 600 ps.


\end{thebibliography}
\end{document}